\documentclass[traditabstract]{aa} 
\usepackage{graphicx}
\usepackage{txfonts}

\usepackage{natbib} 
\bibpunct{(}{)}{;}{a}{}{,} 
\usepackage{url}
\usepackage{multirow,bigdelim}

\def \th {\thinspace}
\newcommand{\sou}{\mbox{NGC 5548 \th}}

\def \kms {\hbox{km s$^{-1}$}}

\def \ergsc{\hbox{erg s$^{-1}$ cm$^{-2}$}}
\def \ergsca{\hbox{erg s$^{-1}$ cm$^{-2}$ A$^{-1}$}}

\def \ergs{\hbox{erg s$^{-1}$}}

\def \msun {\hbox{${\rm M_\odot}$}}

\newcommand\omegam{\hbox{{$\Omega_{\rm m}$}}}
\newcommand\omegalambda{\hbox{{$\Omega_{\Lambda}$}}}
\newcommand\kmsmpc{{\rm km s$^{-1}$ Mpc$^{-1}$}}
\newcommand\ho{\hbox{{$H_{0}$}}}
\def \nh {\hbox{ $N{\rm _H}$ }}
\def \colc {cm$^{-2}$}

\def \xic {erg cm s$^{-1}$}
\def \fwhm {{\em FWHM}}
\newcommand{ \lia} {Ly$ \rm{\,\sc{\alpha}}$}
\newcommand{ \fek} {Fe K$\alpha$}
\def \nhw {\hbox{ $N{\rm _{H,\,  warm}}$ }}
\def \nhc {\hbox{ $N{\rm _{H,\,  cold}}$ }}
\def \cvw {\hbox{ $C{\rm _{V, \, warm}}$ }}
\def \cvc {\hbox{ $C{\rm _{V, \, cold}}$ }}
\newcommand{\xmm }{{\rm XMM-\it{Newton}}}
\newcommand{\chandra }{{\it Chandra\th \th}} 
\newcommand{\leg }{{\it Chandra-\rm{LETGS} \th}} 

\begin{document}

\title{Anatomy of the AGN in NGC 5548}
\subtitle{IV. The short-term variability of the outflows }

\author{L. Di Gesu \inst{1}
       \and E. Costantini \inst{1}
       \and J. Ebrero\inst{2}
       \and M. Mehdipour\inst{1}
       \and J.S. Kaastra\inst{1,3}
       \and F. Ursini\inst{4,5,6}
       \and P.O. Petrucci\inst{4,5}
       \and M. Cappi\inst{7}
       \and G.A. Kriss\inst{8,9}
      \and S. Bianchi\inst{6}
      \and G. Branduardi-Raymont\inst{10}
      \and B. De Marco\inst{11}       
       \and A. De Rosa\inst{12}  
      \and S. Kaspi\inst{13}
      \and S. Paltani\inst{14}
      \and C. Pinto\inst{15}
      \and G. Ponti\inst{11}
      \and K.C. Steenbrugge\inst{16,17}
     \and M. Whewell\inst{10}
}

\institute{
SRON Netherlands Institute for Space Research, Sorbonnelaan 2, 3584 CA Utrecht, The Netherlands \email{L.di.Gesu@sron.nl}
\and XMM-Newton Science Operations Centre, ESA, PO Box 78, 28691 Villanueva de la Canada, Madrid, Spain   
\and Leiden Observatory, Leiden University, Post Office Box 9513, 2300, RA Leiden, The Netherlands
\and Univ. Grenoble Alpes, IPAG, F-38000 Grenoble, France
\and CNRS, IPAG, F-38000 Grenoble, France
\and Dipartimento di Matematica e Fisica, Universit\`a degli Studi Roma Tre, via della Vasca Navale 84, 00146 Roma, Italy
\and INAF-IASF, Bologna, Via Gobetti 101, I-40129, Bologna, Italy
\and Space Telescope Science Institute, 3700 San Martin Drive, Baltimore, MD 21218, USA
\and Department of Physics and Astronomy, The John Hopkins University, Baltimore, MD 21218, USA
\and Mullard Space Science Laboratory, University College London, Holmbury St. Mary, Dorking, Surrey, RH5 6NT, UK
\and Max-Planck-Institute fur extraterrestrische Physik, Giessebachstrasse, D-85748, Garching, Germany
\and INAF/IAPS - Via Fosso del Cavaliere 100, I-00133 Roma, Italy.
\and Department of Physics, Technion-Israel Institute of Technology, 32000 Haifa, Israel
\and Department of Astronomy, University of Geneva, 16 Ch. d'Ecogia, 1290, Versoix, Switzerland
\and Institute for Astronomy, University of Cambridge, Madingley Rd, Cambridge, CB3 0HA
\and Instituto de Astronomia, Universidad Catolica del Norte, Avenida Angamos 0610, Casilla 1280, Antofagasta, Chile
\and Department of Physics,University of Oxford, Keble Road, Oxford, OX1 3RH, UK}

\abstract{
During an extensive multiwavelength 
campaign that we performed in 2013-14 
the prototypical Seyfert 1 galaxy \sou
has been found in an unusual
condition of heavy and persistent obscuration.
The newly discovered ``obscurer'' 
absorbs most of the soft X-ray continuum
along our line of sight and lowers the ionizing
luminosity received by the classical warm absorber.
Here we present the analysis 
of the high resolution X-ray spectra 
collected with \th \xmm \th \th
and \chandra throughout the campaign, which are suitable
to investigate the variability
of both the obscurer and the classical
warm absorber.
The time separation between these X-ray
observations range from 2 days to 8 months.
On these timescales
the obscurer is variable
both in column density 
and in covering fraction.
This is consistent with 
the picture of a patchy wind.
The most significant variation 
occurred in September 2013 
when the source 
brightened for two weeks. 
A higher and steeper
intrinsic continuum 
and a lower obscurer 
covering fraction
are both required 
to explain 
the spectral shape during
the flare. We suggest
that a geometrical change
of the soft X-ray source 
behind the obscurer cause 
the observed drop 
in the covering fraction.
Due to the higher soft X-ray continuum level
the September 2013 \chandra spectrum 
is the only X ray spectrum of the
campaign where individual 
features of the warm absorber 
could be detected.
The spectrum shows 
absorption from  Fe-UTA,
\ion{O}{iv}, and \ion{O}{v},
consistent to belong
to the lower-ionization
counterpart of the historical
\sou warm absorber.
Hence, 
we confirm that the 
warm absorber
has responded to the 
drop in
the ionizing luminosity
caused by the obscurer.}

\titlerunning{Anatomy of an AGN in NGC 5548}
\authorrunning{L. Di Gesu et al}
 \maketitle


\section{Introduction}
\label{intro4}
%
\begin{figure}[t]
    \includegraphics[width=0.5\textwidth]{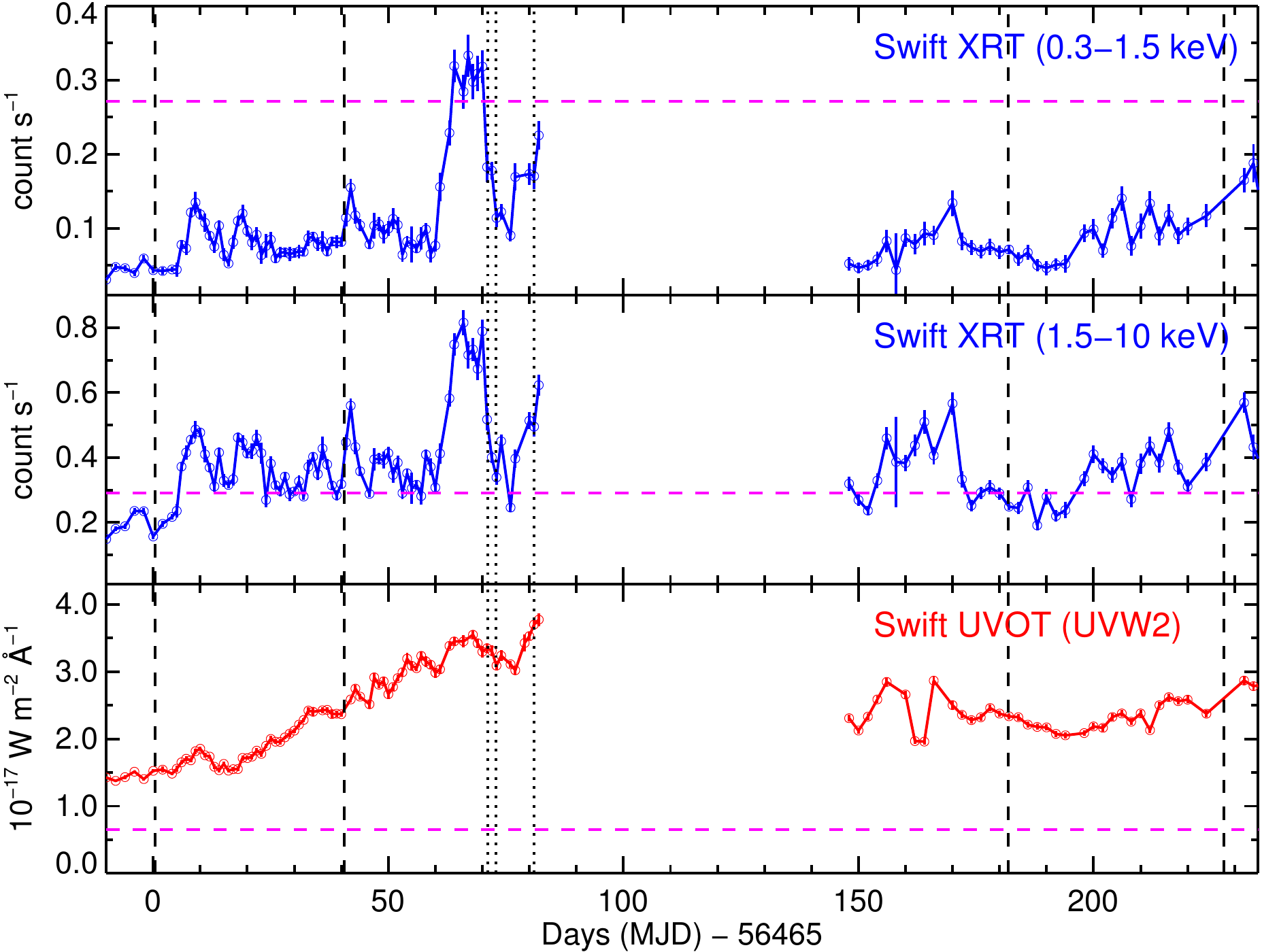}
  \caption{From the top to the bottom panel:
  the observed light curves of \sou in two X-ray bands 
  and in the UV ($\lambda$=2030 \AA). 
  These curves are obtained from the daily Swift monitoring performed during our campaign.
  In each panel, the pink horizontal dashed line marks
  the flux level measured by Swift at unobscured epochs (2005 and 2007). 
  From the left to the right, the vertical lines indicate the first
  and the last \th \th \xmm \th observation of summer 2013
  (dashed lines), the three \chandra observations of September
  2013 (dotted lines), and the last two observations of the
  \th \xmm \th \th program (in December 2013 and February 2014, dashed lines).
  }
  \label{lc.fig}
\end{figure}

%
%
%
In the X-ray band, 
active galactic nuclei (AGN) 
are variable emitters.
The origin of this variability,
that can be even large and fast
\citep[e.g., ][]{mat2003},
is not fully understood yet
\citep[see e.g. ][]{mch2006,tur2009,pon2012}. 
Variable absorption of 
the X-ray radiation along 
the line of sight
is one of the possible explanations.
Indeed, many absorbing components,
spanning a broad range in ionization,
can be detected in AGN X-ray spectra.\\
Cold neutral absorption
is able to strongly suppress 
the soft X-ray flux
and to change the curvature 
of the X-ray 
spectrum.
In recent years,
evidence for variability
due to cold X-ray absorption
in both type 1 and type 2 AGN
has increased.
In many cases
changes in the absorber
column density and/or
covering fraction
on a few hours--few years timescale
have been reported
e.g., in NGC 4388 \citep{elv2004},
NGC 4151 \citep{puc2007}, and
1H 0557-385 \citep{cof2014}.
On these same timescales,
the X-ray absorbing column density  
may even drastically switch 
from Compton thick to Compton thin
as observed in several type 2 AGN
e.g.
NGC 7582 \citep{pic2007},
UGC 4203 \citep{ris2010}, 
and NGC 454 \citep{mar2012}.
This rich phenomenology
suggests that
cold gas
is present even in the
innermost region of AGN
\citep[see][]{bia2012}.
This material is probably
patchy and may belong
to the Broad Line Region
(BLR) or to a clumpy torus
\citep[e.g.,][]{min2014}.
In the best studied
Seyfert 1.8 source NGC 1365
a single cloud has been monitored
while eclipsing the central
X-ray source for a few hours
\citep{ris2009}. In the last
decade, this source
has been observed in a 
vast variety of spectral 
appearances, 
ranging from Compton thick
\citep{ris2005}
to an almost unobscured condition
more reminiscent 
of a pure type 1 AGN
\citep{wal2014}.\\ 
Absorbers at higher ionization
(the so-called warm absorbers, WA)
imprint discrete absorption lines 
on $\sim$50\% of type 1
AGN spectra \citep{cre2003}. 
These features, 
that fall mainly
in the soft X-ray (e.g. 
from 
\ion{Ne}{ii}--\ion{Ne}{x},
\ion{Fe}{x}--\ion{Fe}{xxiv},
and \ion{O}{iv}--\ion{O}{vii}) 
and in the
UV (e.g., \ion{O}{vi}, \ion{Mg}{ii}, 
\ion{C}{ii}--\ion{C}{iv},
and \ion{Si}{ii}--\ion{Si}{iv}) 
domain,
are usually
blueshifted with 
respect to the systemic 
velocity, indicating therefore 
a global outflow of the  absorbing
gas \citep{cre2003}.
In local Seyfert 1 
galaxies
absorption lines
have usually a narrow
profile (\fwhm $\sim10^2$ \kms).
In about 15\% of optically
selected quasars
\citep{wey1991}
broad absorption lines 
(BAL) with a typical width
of $10^4$ \kms \th 
have been also observed
\citep{ham2004}.
In the last ten years
observational campaigns 
providing simultaneous 
high resolution UV 
and X-ray spectroscopy
have been performed
for a handful 
of local AGN
\citep[see][for a review]{cos2010}.
In some cases
(e.g.,
NGC 4151, \citet{kra2005}
Mrk 279, \citet{ara2007,cos2007}
1H 0419-577 \citet{dig2013}), 
it has been
established
that X-ray and UV 
narrow absorption lines (NAL)
may be the manifestation
of the same outflow,
even though it is not 
trivial to determine
how the X-ray and UV
absorbing gas are physically
and geometrically related.
The UV and X-ray absorbing gas
could be co-located
\citep[e.g., Mrk 509, ][]{ebr2011},
with the UV absorbing components
being possibly denser clumps 
embedded in a more highly ionized
wind \citep[e.g., NGC 4051, ][]{kra2012}.
WAs are usually 
complex multicomponent
winds spanning a broad range 
in velocity and in ionization.
The lower ionization
components produce the UV lines
while the higher ionization phases
are seen only in the X-rays.
An intermediate phase
producing absorption lines 
in both bands
may in some cases be present.\\
Variability 
in the warm absorption
may in principle contribute to
the overall 
X-ray variability of AGN
on different timescales.
In order
to assess the 
WA variability on long 
(e.g., $\sim$years) 
timescales 
it is necessary that
high quality multiepoch
spectroscopy is available,
which is seldom the case.
A multiepoch study of the
WA has been attempted
for instance in Mrk 279
\citep{ebr2010}
without finding 
significant variability.
In a different case
\citep[Mrk 841,][]{lon2010},
comparing two different
observations taken $\sim$4
years apart,
a moderate 
decrease 
in the WA ionization as
the continuum dims
has been observed.
An even more noticeable
long-term WA variability
has been reported in the case
of Mrk 335 \citep{lon2013},
where the emergence of an ionized
outflow that was not
historically present
has been observed
in 2009. 
On shorter timescales,
changes
in the WA opacity
or ionization
have been
observed for instance in
NGC 3783 \citep[$\sim$31 days,][]{kro2005}
and
NGC 4051 \citep[few ks--few months,][]{kro2007,ste2009}.\\
The timescale
over which absorption lines
are observed to vary
can be used
to measure the distance
of the absorbing gas
from the ionizing source
\citep[see][]{cre2003}.
Indeed,
for photoionized gas
in equilibrium,
the recombination
timescale
depends on
the gas number density $n$.
Besides the distance $r$
it is the only 
other unknown parameter 
in the definition of 
ionization parameter
$\xi=L_{\rm ion}/{nr^2}$
(where $L_{\rm ion}$ 
is the ionizing luminosity
in the 1--1000 Ryd band).
Searching for
absorption line variability
on short timescales
and thus constraining the
location of the WA
is the main motivation 
for conducting
monitoring 
campaigns of AGN
\citep
[e.g., Mrk 509, 
see][]{kaa2012}.
The knowledge of the
location is crucial
for estimating the mass
outflow rates and kinetic
luminosities associated
to these absorbers 
\citep[e.g.,][]{cre2012}
and thus, 
to evaluate their potential
impact on the host galaxy 
environment.\\
%
%
%
%
%
%

\begin{table*}[t]
\caption{List of the \xmm \th \th 
and \chandra datasets used for the present analysis.}
\centering
\begin{tabular}{lccccc}
\hline
\hline
Obs. & 
Date &
Satellites &
$F_{\rm 2030 \, \AA}$ \tablefootmark{a} 
& $F_{\rm 0.3-2.0\, keV}$ \tablefootmark{b} 
& $F_{\rm 2.0-10.0\, keV}$ \tablefootmark{b} \\
& yyyy-mm-dd & &
($ 10^{-14} \ergsca$) & \multicolumn{2}{c}{($ 10^{-12} \ergsc $)} \\
\hline
XM1 &
2013-06-22&
1 4&  
2.07 &1.1 & 13.2 \\
XM2 &
2013-06-30 &
1 4&  
2.28 & 3.5 & 29.9 \\
XM3 &
2013-07-07 &
1 4&  
2.07 & 2.1 & 21.2 \\
XM4 &
2013-07-11 &
1 4&  
2.32 & 3.6 & 32.1 \\
XM5 &
2013-07-15 &
1 4&  
2.57 & 2.6 & 26.9 \\
XM6 &
2013-07-19 &
1 4&  
2.74 & 2.3 & 27.4 \\
XM7 &
2013-07-21 &
1 4&  
2.88 & 2.3 & 23.5 \\
XM8 &
2013-07-23 &
1 4&  
3.09 & 2.2 & 25.1 \\
XM9 &
2013-07-25 &
1 4&  
2.97 & 3.2 & 29.8 \\
XM10 &
2013-07-27 &
1 4&  
3.29 & 3.0 & 29.4 \\
XM11&
2013-07-29 &
1 4&  
3.21 & 2.8 & 26.4 \\
XM12 &
2013-07-31 &
1 4&  
3.21 & 2.2 & 23.6 \\
\hline \hline
CH1 &
2013-09-01 &
2 4 & 
4.54 & 7.7 & 38.7 \\
CH2 &
2013-09-02 &
2 4& 
4.48 & 4.4 & 29.7 \\
CH3 &
2013-09-10 &
2 3 4& 
4.89 & 5.7 & 36.1 \\
\hline \hline
XM13 &
2013-12-20 &
1 4&  
3.27 & 2.1 & 21.8\\
XM14 &
2014-02-05 &
1 4&  
3.49 & 4.1 & 24.4 \\
\hline
\end{tabular}
 \tablefoot{
 Instruments are labeled with numbers, as followed:
 1: \xmm, 2: \textit{Chandra}, 3: NUSTAR, and 4: Swift.
\tablefoottext{a}{UV flux measured by Swift-UVOT, corrected for reddening.}
\tablefoottext{b}{Observed flux in the quoted bands, as derived from our best fit model.}}
\label{obs4.tab}
\end{table*}

%
%
%
%
With this aim,
during the summer of 2013 
and the winter of 2013-14
we performed
a large multiwavelength 
monitoring campaign on the
bright Seyfert 1 NGC 5548.
The overview of the campaign in
presented in
\citet[][hereafter Paper I]{meh2015}.
NGC 5548
is a prototypical
 Seyfert 1 galaxy,
that has been studied
for decades
 from optical
\citep[e.g.,][]{pet1999}
 to X-ray wavelengths
 \citep[e.g.,][]{nan1993, iwa1999}.
From a dynamical modeling
of the BLR \citep{pan2014},
it is inferred that
this source is observed
at an inclination angle of $\sim30^{\circ}$
and hosts a
supermassive black hole (SMBH)
of $\sim 3 \times 10^7$ \msun
\th in its center.
Previously,
high resolution
UV
\citep{cre1999,cre2003a}
and X-ray 
\citep{kaa2000, kaa2002,ste2003,ste2005}
spectra
have revealed
several deep NAL
that can be ascribed
to a moderate
velocity 
($ v_{\rm out}$=200--1200 \kms) ionized
outflow. \\
%
%
%
Unexpectedly, 
throughout 
the whole 2013-14 campaign
\sou appeared
dramatically different
from the past
\citep[e.g., from the \chandra observation of 2002,][]{ste2005}
being $\sim$25 times
less luminous in the soft X-rays.
Moreover,
it showed
broad, asymmetric absorption troughs
in the blue wings of
the main UV broad
emission lines
(e.g., in
\lia, \ion{C}{iv}, \ion{N}{v}).
In \citet[][ hereafter K14]{kaa2014}
we proposed
that all these changes 
can be ascribed to
the onset of 
a persistent,
weakly ionized,
fast 
(v$\sim$5000 \kms),
wind
(hereafter ``the obscurer'').
The obscurer is located within 
or just outside the BLR, 
at a distance of a few light 
days from the SMBH, 
and possibly
has been launched 
from the accretion disk. 
It blocks 
$\sim$90\% 
of the
X-ray flux along
our line of sight,
thereby lowering
the ionizing
luminosity
received by
the WA.
Indeed, in this
obscured condition,
the historical
\sou WA is
still present, but
with a lower ionization.
In the X-rays it is consistent
with being $\sim$3 times
less ionized  than
what was observed in 2002 (K14),
and, at the same time,
in the UV it shows
new lower-ionization
NAL
\citep[from e.g. \ion{C}{ii}
and \ion{C}{iii},][in press]{ara2015}.\\
In this paper
we use all
the 
high resolution 
X-ray spectra 
collected during our
campaign
to
assess 
what drives
the spectral changes
of
\sou on timescale
as short as few days.
This is the typical
time separation between
the X-ray observations
of the campaign.
These spectra are
suitable to investigate
the absorption variability,
because they 
cover the energy band
where the main ionized
and neutral absorption
features of e.g.
oxygen and iron fall.
During the campaign
the source 
was always weak
in the soft X-rays,
except 
for a sudden brightening
in September 2013 
(Fig. \ref{lc.fig}). 
On that occasion,
we triggered
a \leg observation.
In the following we investigate
also the possible causes and
consequences of this
sudden brightening.\\
The paper is organized as follows:
in Sect. \ref{data4} we briefly
present the datasets that we use
in this analysis, and
in Sect. \ref{templ} we describe the
template spectral model that we apply
to all the datasets in Sect. \ref{fits}.
Finally in Sect. \ref{disc4} 
we discuss our results
and in Sect. \ref{conc4} 
we outline the conclusions.\\
The C-statistic \citep{cas1979} is used
throughout the paper,
and errors are quoted at 68$\%$ 
confidence level
($\Delta C=1.0$). In all the
spectral models presented in the
following, we use the Galactic 
hydrogen column density
from \citet[][\nh=$1.45 \times 10^{20}$ \colc]{wak2011}.
The cosmological redshift
that we adopted for \sou is 0.017175
\citep{dev1991}.The cosmological parameters are set to:
\ho=70 \kmsmpc, \omegam=0.3 and \omegalambda=0.7.


\section{The data}
\label{data4}
\xmm \th observed \sou 
between June 2013 and February 2014
using both the EPIC cameras 
\citep{tur2001,stru2001} and
the Reflection Grating Spectrometer
\citep[RGS,][]{den2001}.
The core of the
campaign 
consisted of 12,
$\sim$50 ks long, \xmm \th \th
observations that were
taken every $\sim$ 2--8 days 
in June and July 2013.
After these, two other observations
were acquired in December 2013
and February 2014, providing
14 \xmm \th \th datasets
in total.
The details of all
the observations,
that hereafter 
we label 
in chronological
order as XM1--XM14,
are given in Table
\ref{obs4.tab}.
During the entire
campaign, the source was also 
monitored daily
by Swift \citep{geh2004},
both in the X-rays
with the X Ray Telescope 
\citep[XRT,][]{bur2005}
and in the UV with the
UV Optical Telescope 
\citep[UVOT,][]{rom2005}.
Swift-UVOT flux measurements 
in the UVW2 
($\lambda$=2030 \AA) filter,
corrected for reddening and
for the host galaxy contribution
as explained in Paper I, 
are also used in the 
present analysis.\\
In September 2013
we triggered a \chandra 
observation
because
Swift observed a sudden
brightening (Fig. \ref{lc.fig}).
The observation was performed
with the Low Energy Transmission
Grating Spectrometer \citep[LETGS,][]{bri2000}
in combination
with the High Resolution
Camera (HRC-S).
The observing time
was split in three observations
(Obs. CH1--CH3)
of 30, 67 and 123 ks respectively.
The first two were
taken on September 1st and 2nd
while the third and longest one
was taken a week later on September
10th.
Simultaneously with this observation
a higher energy spectrum was
acquired with the
Nuclear Spectroscopic Telescope Array
\citep[NuSTAR,][]{har2013} satellite.
In the occasion of this flaring event,
no \xmm \th \th observation
was available.\\
A detailed description
of the data reduction
procedure for all the instruments
is given in Paper I.
In the present analysis,
we fit the high resolution
RGS and
LETGS spectra.
The simultaneous EPIC-pn
and NuSTAR spectra
at higher,
less absorbed, energies
provide the continuum
baseline over which the
absorption is superposed.
We defer the reader
to \citet[][in press]{urs2015},
hereafter U15, 
for a detailed
modeling of the continuum
at high energies.\\
As in K14,
we fitted 
the EPIC-pn in 
the 1.03--10.0 \AA\
($\simeq$1.24--12 keV) band
together with the
RGS in the 
5.68--38.23 \AA\
($\simeq$0.32--2.2 keV)
band. Thus, the two
instruments overlap
in a small band
allowing to check 
for possible
intercalibration
mismatches.
In all our fits
the intercalibration
factor used in K14
($\sim$ 1.027) was
adequate.
Due to an incomplete 
correction 
for the gain of the EPIC-pn
(not corrected in the
SAS v13 we used, see Paper I), 
the \fek \th line appears blueshifted 
(see Cappi et al. 
in prep. for a detailed discussion), 
which we correct for with 
an artificial redshift 
for these spectra. 
However, 
this solution leads  
to a poor fit near the energy
of the gold M-edge of the telescope
mirror.
For this reason, 
we omitted
the interval 5.0--6.2 \AA\
($\simeq$2.0--2.5 keV)
from all our fits.\\
We fitted the 
\leg spectra 
between 2 and 40 \AA\
($\simeq$0.3--6.2 keV).
We used the NuSTAR spectrum
simultaneous to CH3
in the 0.2--2.5 \AA\
($\simeq$5--60 keV) band.
Since \leg and NuSTAR 
were consistent
in the overlapping 
band we did not
apply any 
intercalibration correction
in the joint fit.


\section{The template model}
\label{templ}
In order to assess 
the variability of NGC 5548, 
we used the same
template model to fit all the spectra.
This template is
close to the model adopted in K14,
differing from it only in 
the modeling
of the ``soft-excess'' 
\citep{arn1985} component.
We performed 
all the spectral
analysis
with the latest version of SPEX
\citep[v. 2.05,][]{kaa1996}.
In our template model
we considered
the cosmological redshift
and the Galactic absorption.
For the latter,
we use a collisionally ionized
plasma model (HOT), with a nominal
temperature of 0.5 eV for a neutral
gas.\\
Our continuum model includes 
a primary and reflected power-law, 
on top of which lies
a soft-excess component.
In \sou the reflection
is consistent to be constant
(see U15) thus producing
a steady narrow \th \fek \th line 
(see Cappi et al., in prep.).
In the present analysis
we used the reflection
parameters obtained in U15
($\Gamma$=1.9 and $E_{\rm cut}$=300 keV
for the photon index and the high energy
cutoff of the primary power law)
in the SPEX reflection model
REFL, which includes both a
Compton reflected continuum
\citep{mag1995}
and the \fek \th line
\citep{zyc1994}.
To adjust the 
fit of the \fek \th \th
line, we let the normalization
of REFL free to vary
within the errors of the
U15 model.
When using EPIC-pn data,
we applied to REFL an artificial blueshift
(z=--1.25$\times 10^{-2}$) 
to correct
for an apparent centroid shift 
of the \th \fek \th line
(see also Sect. \ref{data4}).\\
%
%
%
%
%
%
The X-ray obscuration affects 
the spectrum mainly below 2.0 keV
and makes therefore the detection
of the soft-excess 
from X-ray spectra elusive. 
Nevertheless,
this component
contributes to the continuum
in the band where most
of the absorption is
seen, thus
its modeling
is critical 
for evaluating
the absorption 
variability.
In Paper I we show that 
the soft-excess component is 
likely to be the tail
of a component extending from UV to soft X-rays, 
which is produced 
by Compton up-scattering of the disk photons
in a warm, optically thick plasma. 
Therefore, in this framework,
the UV emission (which is not affected by the obscuration) 
is a proxy for the X-ray soft-excess. 
We used COMT \citep[based on][]{tit1994}
to model the soft-excess component 
and we used the UV flux ($F_{\rm 2030 \, \AA}$) 
listed in Table \ref{obs4.tab} 
for each observation to set the normalization.
The 0.3--2.0 keV luminosity 
of the soft-excess is given by: 
$(L^{\rm COMT}_{\rm 0.3-2.0\,keV}/10^{34} {\rm W}) 
=2.093+2.893 \times (F_{\rm 2030 \, \AA}/10^{-34} \rm Jy)$, 
valid for the range of observed UV fluxes used here
(Mehdipour, private communication).
Hence, consistently with the 
long-term variability analysis 
in Mehdipour et al. (in prep.), 
we kept the COMT component
constant in shape 
in all the fits,
with its 0.3-2.0 keV 
flux varying according to the UV flux.
We use the mean values 
given in Paper I 
for the other COMT parameters,
namely
the Wien temperature 
of the incoming photons 
($T_{\rm{0}}=80$ eV),
the temperature 
($T_{\rm{1}}=0.17$ keV), 
and the optical depth 
($\tau=21$) of the plasma.\\
A detailed
modeling of the X-ray emission features
of \sou will be presented in Whewell et al,
in preparation.
The X-ray narrow
emission lines
are  consistent with being 
constant during the
\xmm \th campaign
and thus could be kept
frozen in all our fits.
For the narrow lines
we the values
obtained in the
K14 fit,
while for the
broad emission features we
used the fluxes given in
\citet{ste2005}.\\
In K14 we found that
the obscurer
causing the persistent flux dimming
of  \sou
comprises two ionization phases
(hereafter labeled as ``warm''
and ``cold").
The warm phase is
mildly ionized ($\log \xi=-1.25$) 
and has a larger covering 
fraction 
(e.g., $\cvw \sim 86\%$, 
$\cvc \sim 30 \%$, K14) 
and a lower column density 
(e.g., $\nhw \sim 10^{22}$ \colc, 
$\nhc \sim 10^{23}$ \colc, K14)
than the cold phase
which is consistent with being neutral. 
In SPEX, we used 
a photoionized absorber model (XABS)
for both the obscurer components
and we adopted for them
the same kinematics
(outflow and broadening velocity) 
used in K14. 
The ionization
balance for both the obscurer
components is computed using
the intrinsic, unabsorbed spectral
energy distribution (SED).\\
Finally,
we included six more XABS components
in the template model to account
for the lower ionized counterpart
of the historical \sou warm absorber.
The ionization balance
for all the WA components
is computed using
a SED filtered 
by the obscurer.
We label the WA components with
capital letters, from A to F
as the ionization increases.
As in K14,
we assumed that
the WA varies 
only in ionization
when the ionizing SED
illuminating it changes.
Indeed,
from UV spectra 
we know that the kinematics 
of the WA components has not changed 
over at least 16 years, 
and Ebrero et al. (in prep.) 
shows no historical evidence 
for total hydrogen column 
density variations.
We kept therefore the kinematics
and the equivalent 
hydrogen column density
of all the WA components
frozen to the parameters
obtained from
the unobscured 2002 spectrum
(see the updated fit 
in Ebrero et al., in prep.). \\
In the following
we apply this template model 
to all the \th \xmm \th \th
and \chandra datasets of
our monitoring campaign.
In all the fits,
the free parameters
are the slope and the normalization
of the primary power law,
the column density, and the
covering fraction of
the two obscurer phases.
Besides these, in the fits 
of those datasets where
either the continuum
or the obscurer parameters
change significantly,
we also allow the ionization 
parameters of the 
warm absorber to vary.
The variability 
in flux of the 
soft excess component
was determined directly
from the Swift UV flux,
as explained above.
%
%
%
%

\section{Fitting the variability}
\label{fits}

\begin{figure}[t]
  \centering
  \begin{minipage}[c]{.50\textwidth}
    \includegraphics[width=1.0\textwidth]{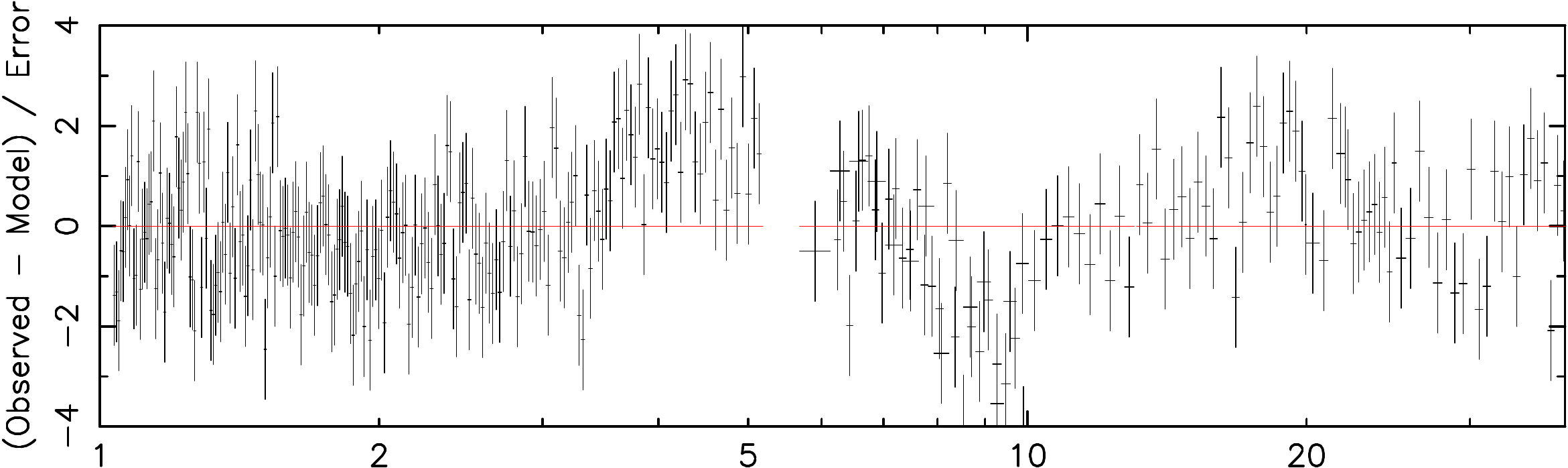}
    \includegraphics[width=1.0\textwidth]{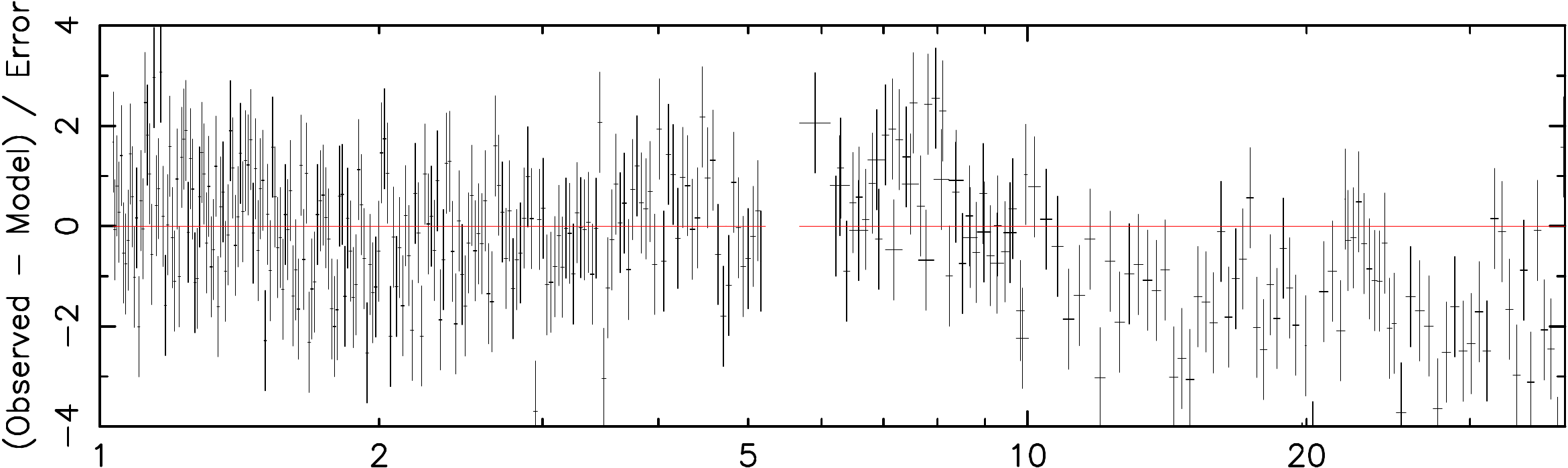}
    \includegraphics[width=1.0\textwidth]{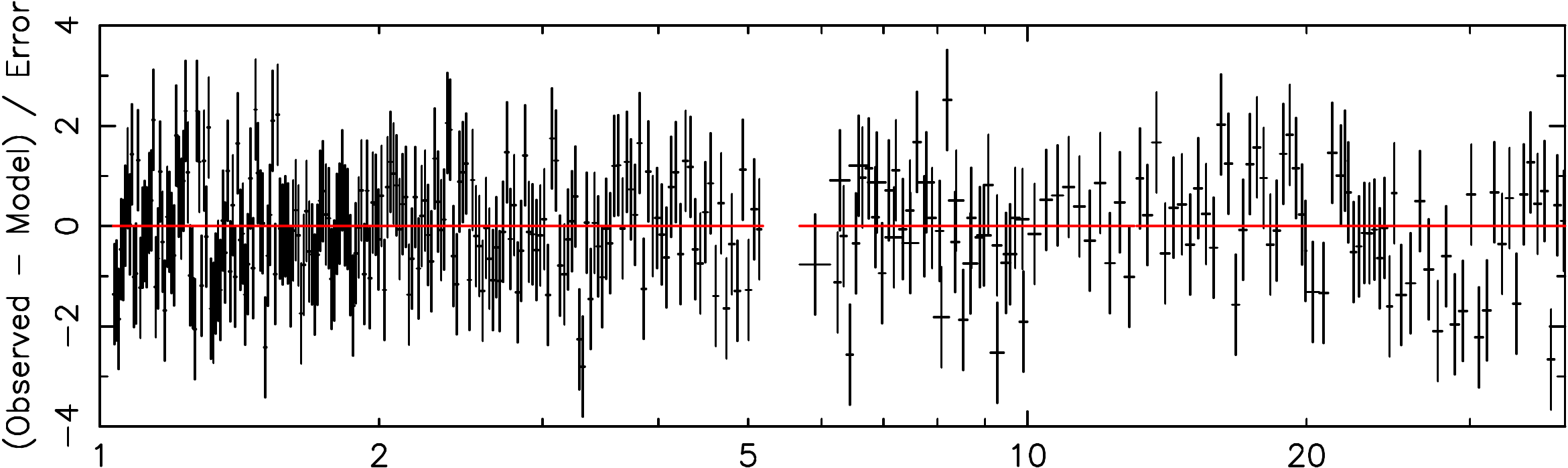}
    \includegraphics[width=1.0\textwidth]{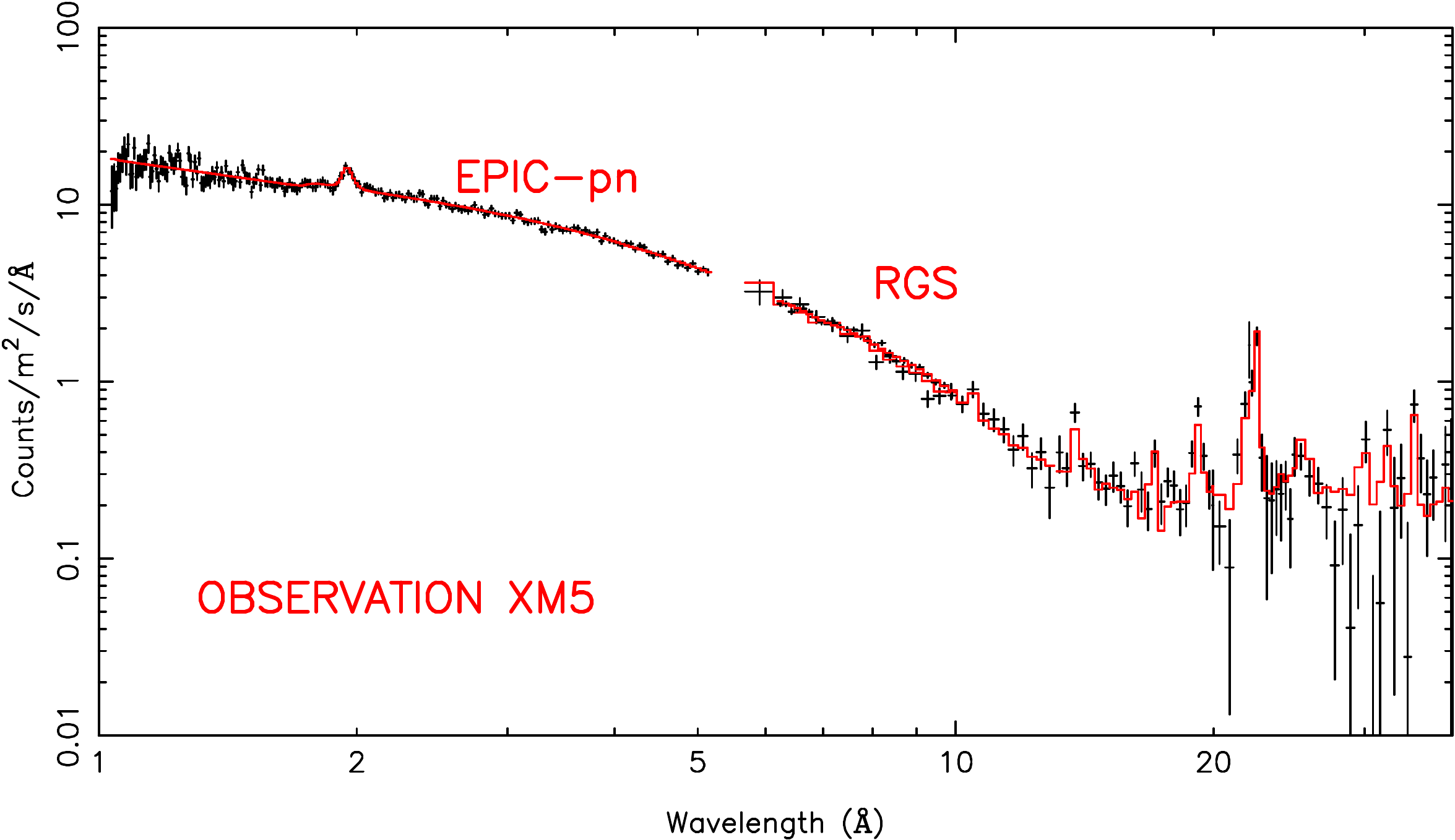}
  \end{minipage}
  \caption{Examples of fits the \xmm \th \th spectrum of Obs. XM5.
  From the top to the bottom panel: 
  fit residuals when only the obscurers covering fractions are allowed to vary 
  from the values obtained for the average spectrum;
  fit residuals when only the obscurers column densities are allowed to vary
  from the values obtained for the average spectrum;
  fit residuals of the best fit model where both column densities 
  and covering fractions are permitted to vary freely;
  spectrum of observation XM5. The solid line represents the best fit model.
  The data have been rebinned for clarity.}
\label{xmspettro.fig}
\end{figure}



\begin{table*}[t]
\caption{Best fit parameters and errors, 
for the individual \th \xmm \th \th and Chandra observations.
In the last row, we list also the parameters derived in K14 for the average XM1--12 spectrum
for comparison.}
\setlength{\tabcolsep}{3pt}
\begin{tabular}{lcccccccccccc}
\hline
\hline
& & \multicolumn{4}{c}{Obscurer components \tablefootmark{a}}
& \multicolumn{6}{c}{WA components \tablefootmark{b}} & \\
%
%
Obs &
$\Gamma $ \tablefootmark{b}&
\nhw   \tablefootmark{c}&
\cvw \tablefootmark{d}&
\nhc \tablefootmark{c}&
\cvc \tablefootmark{d}&
$ \xi_{\rm A}$ \tablefootmark{e}&
$ \xi_{\rm B}$ \tablefootmark{e}&
$ \xi_{\rm C}$ \tablefootmark{e}&
$ \xi_{\rm D}$ \tablefootmark{e}&
$ \xi_{\rm E}$ \tablefootmark{e}&
$ \xi_{\rm F}$ \tablefootmark{e}&
C/Expected C\tablefootmark{f}\\
& & ($10^{22}$ \colc) &
& ($10^{22}$ \colc)
& &
\multicolumn{6}{c}{(log \xic)}\\
\hline
XM1 &
1.5 \tablefootmark{**}&
$ 1.7 \pm 0.1$ & $ 0.86 \pm 0.01$ &
$ 12 \pm 1$ & $ 0.47 \pm 0.02$ &
\rdelim\{{12}{10pt}
\multirow{12}{*}{0.33$^{*}$} &
\multirow{12}{*}{1.06$^{*}$} &
\multirow{12}{*}{1.70$^{*}$} &
\multirow{12}{*}{1.91$^{*}$} &
\multirow{12}{*}{2.48$^{*}$} &
\multirow{12}{*}{2.67$^{*}\,$}
\rdelim\}{12}{10pt} & 384/334\\
XM2 &
$1.58 \pm 0.03$&
$ 1.04 \pm 0.04$ & $ 0.927 \pm 0.005$ &
$ 10 \pm 2$ & $ 0.170 \pm 0.03$ & & & & & & & 422/345 \\
XM3 &
1.5 \tablefootmark{**}&
$ 1.42 \pm 0.06$& $ 0.909 \pm 0.005$ &
$ 11 \pm 3$ & $ 0.18 \pm 0.02$ & & & & & & &379/337 \\
XM4 &
$1.60 \pm {0.03}$&
$ 1.17 \pm 0.04$& $ 0.922 \pm 0.004$ &
$ 14 \pm 2$ & $ 0.19 \pm 0.03$ & & & & & & &406/346 \\
XM5 &
$1.57 \pm 0.03$ &
$ 1.44 \pm 0.05$& $ 0.915 \pm 0.004$ &
$ 8 \pm 1$ & $ 0.24 \pm 0.03$ & & & & & & &435/335 \\
XM6 &
1.5 \tablefootmark{**}&
$ 1.4 \pm 0.1$& $ 0.916 \pm 0.005$ &
$ 5 \pm 1$ & $ 0.30 \pm 0.08$ & & & & & & & 415/341 \\
XM7 &
1.5 \tablefootmark{**}&
$ 1.40 \pm 0.05$& $ 0.909 \pm 0.005$ &
$ 12 \pm 2$ & $ 0.27 \pm 0.03$ & & & & & & & 394/340 \\
XM8 &
$1.53 \pm 0.03$&
$ 1.62 \pm 0.07$& $ 0.88 \pm 0.05$ &
$ 9 \pm 1$ & $ 0.35 \pm 0.03$ & & & & & & & 389/339 \\
XM9 &
$1.64 \pm 0.03$&
$ 1.58 \pm 0.05$& $ 0.896 \pm 0.004$ &
$ 11 \pm 2$ & $ 0.24 \pm 0.03$ && & & & & & 390/340 \\
XM10 &
$1.57 \pm 0.03$&
$ 1.43 \pm 0.05$& $ 0.911 \pm 0.004$ &
$ 9 \pm 1$ & $ 0.24 \pm 0.03$ & & & & & & & 434/344 \\
XM11 &
$1.56 \pm 0.03 $&
$ 1.46 \pm 0.05$& $ 0.902 \pm 0.004$ &
$ 10 \pm 2$ & $ 0.23 \pm 0.03$ & & & & & & & 399/340 \\
XM12 &
1.5 \tablefootmark{**}&
$ 1.43 \pm 0.04$& $ 0.922 \pm 0.005$ &
$ 12 \pm 2$ & $ 0.27 \pm 0.03$ & & & & & & & 447/338 \\
\hline
\hline
CH1 &
1.83 \tablefootmark{*}&
$ 1.7 \pm 0.2$& $ 0.74 \pm 0.03$ &
17 \tablefootmark{*} & $ \leq 0.23 $ &
0.38 \tablefootmark{*} &
1.11 \tablefootmark{*} &
1.75 \tablefootmark{*} &
1.96 \tablefootmark{*} &
2.53 \tablefootmark{*} &
2.72 \tablefootmark{*} & 149/152 \\
CH2+3 &
$ 1.83 \pm 0.02$ &
$ 1.49 \pm 0.08$& $ 0.79 \pm 0.07$ &
$ 17 \pm 3$ & $ 0.35 \pm 0.03$ &
0.38 & 1.11 & 1.75 & 1.96 & 2.53 & 2.72 & 530/311 \\
\hline
\hline
XM13 &
1.5 \tablefootmark{**}&
$ 1.7 \pm 0.1$& $ 0.87 \pm 0.02$ &
$ 9 \pm 2$ & $ 0.33 \pm 0.03$ &
0.33 \tablefootmark{*}& 
1.06 \tablefootmark{*}& 
1.70 \tablefootmark{*}& 
1.91 \tablefootmark{*}& 
2.48 \tablefootmark{*}& 
2.67\tablefootmark{*} 
& 415/340\\
XM14 &
$1.59 \pm 0.01$ &
$1.22 \pm 0.03$ & $ 0.917 \pm 0.003$ &
$10$\tablefootmark{*}  & $ \leq 0.003$ &
0.39 & 1.12 & 1.76 & 1.97 & 2.54 & 2.73 & 485/342 \\
\hline
\hline
K14 \tablefootmark{***} &
$1.566 \pm 0.009$ &
$ 1.21 \pm 0.03$ & $ 0.86 \pm 0.02$ &
$ 9.6 \pm 0.5$ & $ 0.30 \pm 0.10$ &
0.33 &
1.06 &
1.70 &
1.91 &
2.48 &
2.67
& \\
\hline
\end{tabular}
\tablefoot{
\tablefoottext{a}
{For the warm obscurer: $\log \xi=-1.25$. 
For the cold obscurer: $\log \xi=-4$}
\tablefoottext{b}
{For Obs. CH2+3, XM14, and K14
the ionization parameters given are the output 
of the iterative fitting procedure described in Sect. \ref{ch.sec}.}
\tablefoottext{c}
{Photon index of the primary continuum.}
\tablefoottext{d}
{Column density of the obscurer components.}
\tablefoottext{e}
{Covering fraction of the obscurer components.}
\tablefoottext{f}{Ionization parameters of the warm absorber components.}
\tablefoottext{g}{C-statistics of the final best fit model.}
\tablefoottext{*}{Frozen parameters.}
\tablefoottext{**}{Lower limit of the fitting range.}
\tablefoottext{***}{Best fit parameters derived in K14 for the co-added XM1--12 spectrum.}}
\label{allpar.tab}
\end{table*}
%

\subsection{The core of the campaign: from June to August 2013}
\label{chfits}

The \th \xmm \th \th 
monitoring campaign began on June 22nd,
catching \sou at the lowest flux level
observed in any of the high resolution
spectra of the campaign
($F_{\rm 0.3-2.0 \, keV} \sim 1.1 \times 10^{-12}$ \ergsc,
for Obs. XM1, see Table \ref{obs4.tab}).
Throughout all the summer of 2013,
(Obs. XM2--XM12)
the 0.3--2.0 keV flux remained
quite stable (within
a factor $\sim$ 1.3)
around an average level
of $\sim 2.7 \times 10^{-12}$ \ergsc.
The stacked \th \xmm \th \th spectrum 
of Obs. XM1--XM12 
is published in K14.
In that paper
we determine
consistently
both the average obscurer parameters
and the average obscured 
SED illuminating the WA. Accordingly,
we compute the ionization balance
and the ionization parameters
for all the WA components.\\
We fitted Obs. XM1--XM12 assuming
the average WA determined in K14.
Hence, only the continuum
and the two obscurer phases were
left free in the fits.
Furthermore,
in the fits we set a lower limit
for the continuum slope 
($\Gamma \geq 1.5$) to be
consistent with what was observed
during the campaign at higher
energies and with the
analysis repeated by Cappi et al,
in preparation. Indeed,
when fitting such absorbed spectra
below $\sim$10 keV,
a flatter intrinsic continuum
becomes degenerate with a stronger
obscuration. The assumption on the
slope aimed at minimizing this ambiguity
and, however, 
did not affect the
final results
(as we checked 
\textit{a posteriori},
see below).\\
As a first step we fit the spectra 
allowing for both the obscurer phases just one 
of the parameters (column density, 
and covering factor) free.
When only the covering fractions are
allowed to vary the fit displays
strong residuals, for instance
the negative residuals just below $\sim$10 \AA\
(See Fig. \ref{xmspettro.fig}, first panel).
In the observation
where they are more evident (Obs. XM5) 
the C-statistic
is 600 for an expected value of 340.
When we fixed the covering factors 
but instead allowed the column densities to vary, 
more systematic residuals 
between 10 and 20 A are apparent 
(Fig. \ref{xmspettro.fig}, second panel).
Therefore, we
conclude that the two phases of the obscurer
have to be variable both in column density and the
covering fraction to adapt the template model
to all the individual \th \xmm \th \th observations.\\
We tested however the
possibility that the obscurer
varies in ionization rather
than in covering fraction. 
For most of the datasets,
a statistically acceptable
fit can be 
achieved keeping 
the covering fraction
of both the obscurer phases
frozen to the average values
derived in K14
and allowing the ionization
parameter of the warm obscurer
to vary instead.
However,
with these constraints
the best fit
prefers an almost 
neutral obscurer
(e.g. $\log \xi \leq -3.5$),
which would be
too lowly ionized
to produce,
for instance,
the  broad \ion{C}{iv}
absorption lines
that are seen in the UV.
Hence, we discarded these
fits.\\
The final best fit parameters are listed in Table
\ref{allpar.tab}. 
In a couple of cases, the
fit stopped at the lower limit we had imposed
for $\Gamma$. We checked how much further the 
fit of these datasets could 
be improved allowing an even flatter
continuum. In all cases, 
releasing the spectral index 
resulted in a negligible improvement
of the fit (e.g., for Obs. XM1,
$\Delta C=-1$ for $\Gamma$=1.46).\\
The RGS spectra 
of the individual
observations are rather noisy
and the residuals do not
show any hint of 
unaccounted WA features.
However,
as a final test,
we checked
how sensitive
are the best fits 
to possible
variations in the WA
ionization.
We attempted
to refit
each observation
setting
the WA ionization parameters
according to the
variation from the average
of the continuum 
normalization.
All the fits
were insensitive 
to this variation,
(e.g., $\Delta C \sim -1$)
with the free parameters
remaining the same 
within the
errors.
Therefore,
we concluded \textit{a posteriori} 
that assuming a constant WA
in the core of the \th \xmm \th \th
campaign was reasonable.
We show in Fig. \ref{xmspettro.fig} 
(third and fourth panel)
an example of best fit 
(Obs. XM5).
%
%
%
%
%
%

\subsection{The flare of September 2013 as seen with \leg and NuSTAR}
\label{ch.sec}

\begin{figure}[t]
    \includegraphics[angle=180,width=0.5\textwidth]{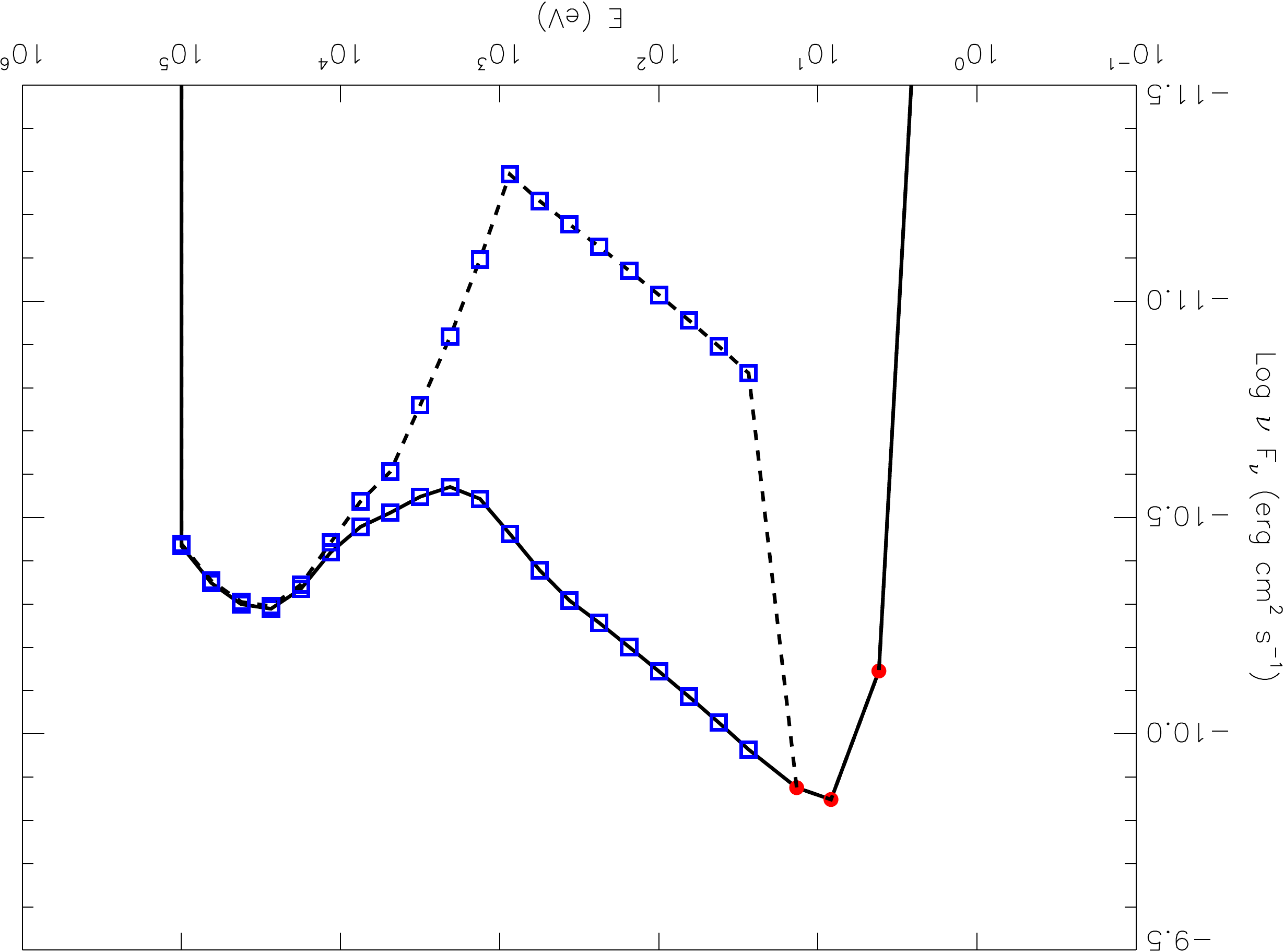}
  \caption{The unobscured (solid line) and obscured (dashed line)
    spectral energy distributions for \sou during the September 2013 flare,
    which were used for the photoionization modeling of the absorbers.
    The 3 data points in the UV (filled circles) are taken from
    the Comptonization model of Paper I. The data points from the EUV to hard X-rays
    (open squares) are also interpolated from the X-ray continuum model.
    See Sect. \ref{chfits} for details.}
  \label{sed.fig}
\end{figure}


\begin{figure}[t]
  \centering
  \begin{minipage}[c]{.50\textwidth}
   \includegraphics[width=1.0\textwidth]{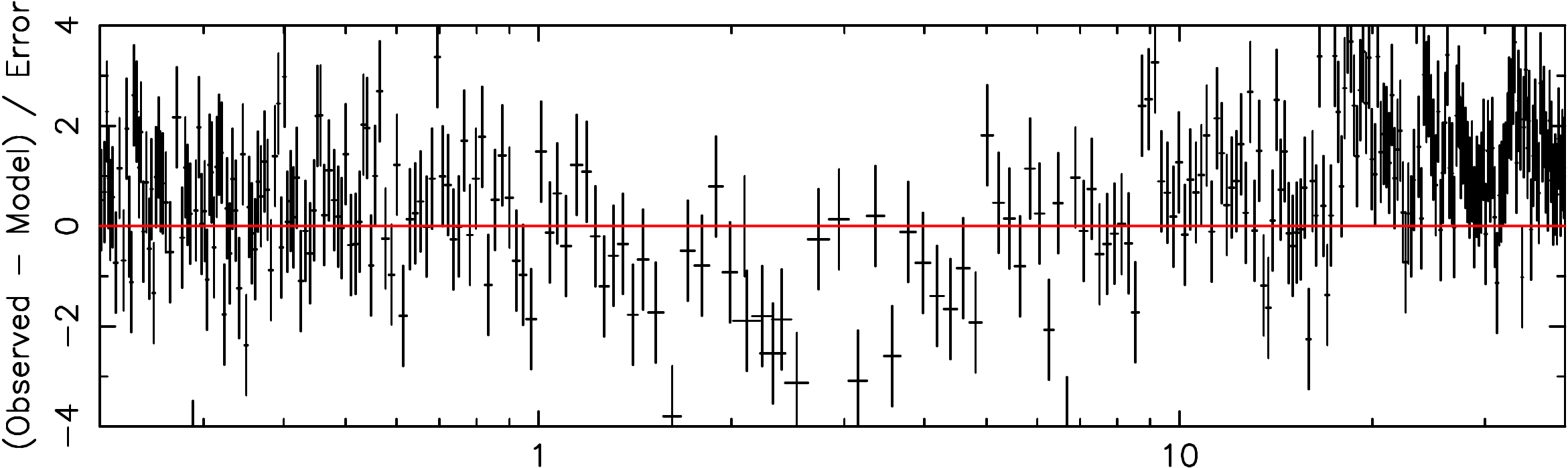}
   \includegraphics[width=1.0\textwidth]{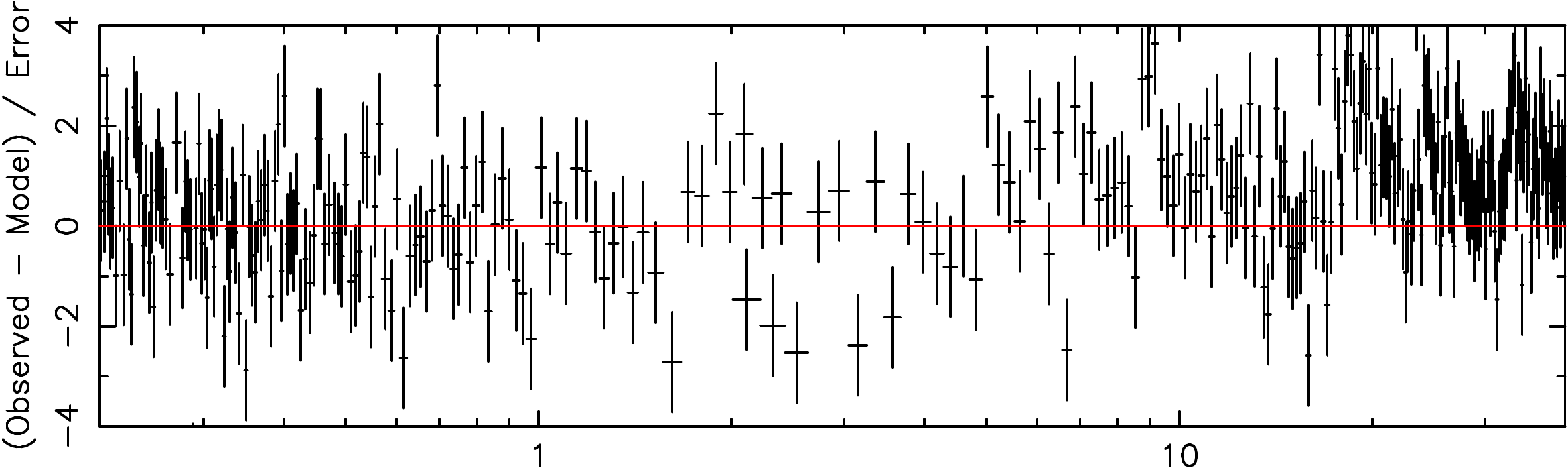}
   \includegraphics[width=1.0\textwidth]{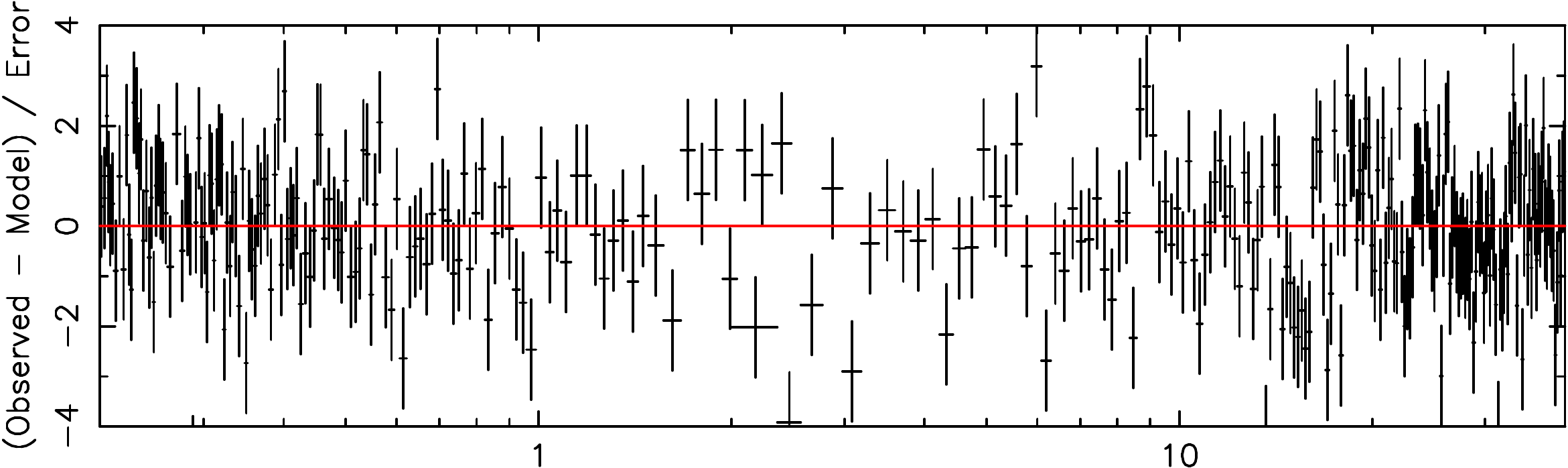}
   \includegraphics[width=1.0\textwidth]{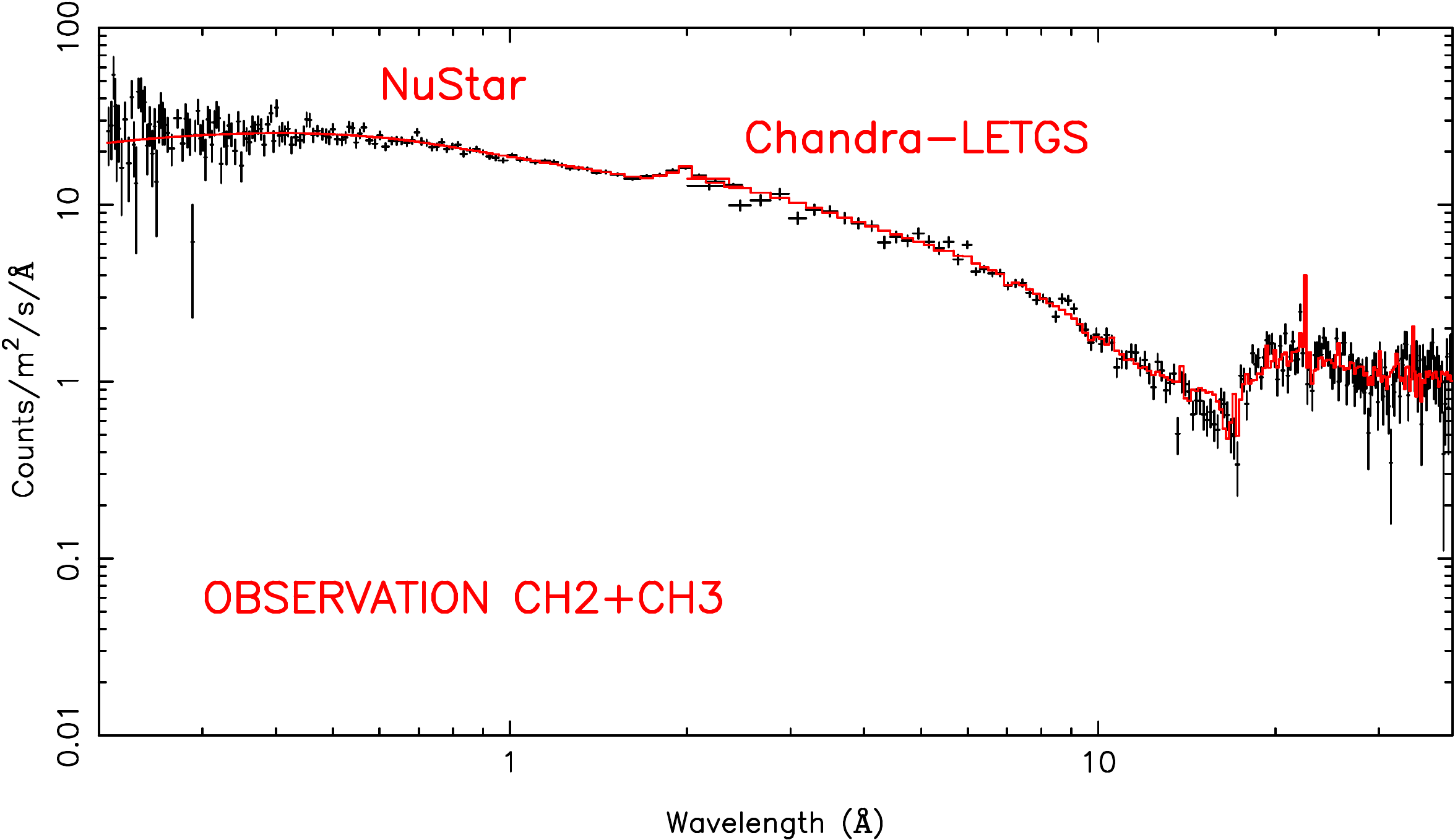}
  \end{minipage}
  \caption{Examples of fit for the \leg plus NuSTAR
            spectrum of \sou during the September 2013 flare.
  From the top to the bottom panel: 
  fit residuals when only the continuum is allowed to vary 
  from the values obtained for the average spectrum;
  fit residuals when only the obscurers covering fraction is allowed to vary
  from the values obtained for the average spectrum;
  fit residuals for the final best fit model. 
  The solid line represent the best fit model.
  The data have been rebinned for clarity.}
\label{chspettro.fig}
\end{figure}


\begin{figure}[t]
    \includegraphics[angle=180,width=0.5\textwidth]{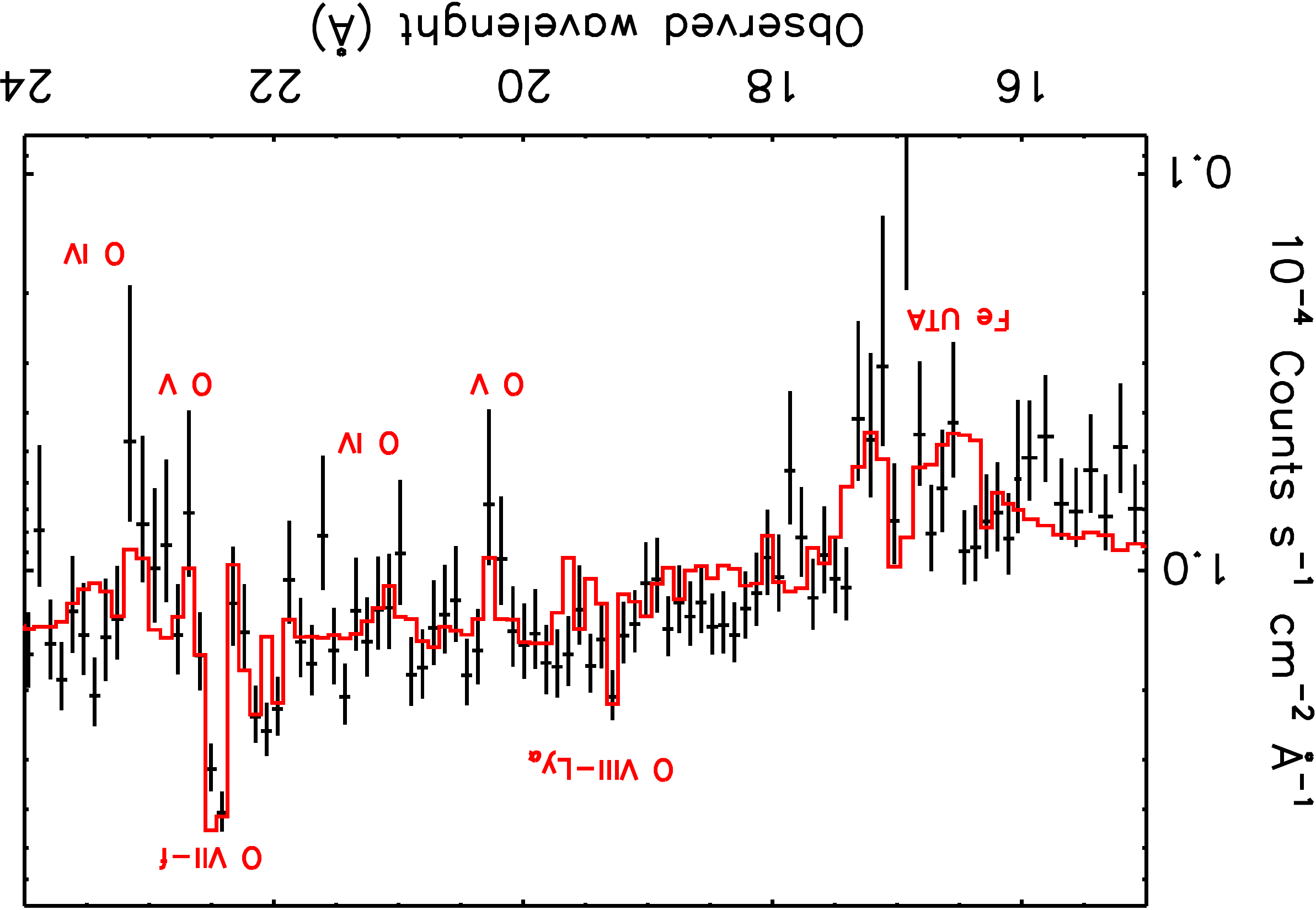}
  \caption{The \leg spectrum of \sou in the 15--24 \AA\ wavelength region.
  The solid line represents our best-fit model. The most prominent emission 
  and absorption features are labeled.
   The spectrum shows 
  some WA signatures (Fe-UTA, \ion{O}{iv}--\ion{O}{v})
  that were not detected during the \xmm \th \th campaign.
  }
\label{lines.fig}
\end{figure}


In September 2013 
we triggered a series of 3 
\leg observations because
it seemed that \sou was recovering
from the obscuration. Indeed,
in a few days, the X-ray flux
in both the Swift-XRT bands rose
above the level
measured at unobscured epochs
and remained steady
for about a week. This brightening was
however a short-lived flare, and after
a few days the source fell again to
the typical low flux level
of the \th \xmm \th \th campaign
(Fig. \ref{lc.fig}).
Our triggered 
\leg observations missed the
peak of the flare. The first 
two observations
were taken during its declining tail, 
while a week later
the third one 
caught a smaller
rebrightening.\\
%
%
%
To understand
if and how
the absorbing components
had responded
to these continuum changes
we needed to use
in the 
photoionization modeling
of the obscurer
a SED
representative 
of \sou during the flare
(Fig. \ref{sed.fig}, solid line).
To construct it,
we used the Comptonization
model of Paper I
which extends
from the UV 
to the soft X-rays.
At the same time,
NuSTAR provided 
the continuum slope
at high energies.
In the UV
we took the model values
corresponding to
$\lambda=2987$ \AA, 
$\lambda=1493$ \AA\
$\lambda=909$ \AA.
We then interpolated 20 data points
in the model between
0.03 and 100 keV.
Finally, we cut off the SED
at low energies 
(below 0.01 Ryd).\\
%
%
%

We derived consistently
the obscurer parameters
and the  WA ionization parameters
using the same 
iterative method of K14.
At each iteration of this
fitting routine
new obscurer parameters
are fitted. Next,
the new obscured continuum
is used as the ionizing SED 
in the photoionization modeling
of the six WA components.
The new ionization
parameters are assigned
by rescaling those observed
in the unobscured spectrum 
of 2002 to the level
of the current obscured continuum.
Explicitly, 
at the N$_{th}$ iteration
$\xi_N/\xi_{2002} 
\sim 
L_{\rm ion, N}/L_{\rm ion, 2002}$.
Finally, the ionization
balances for the new ionization parameters
are recomputed before moving to the
next iteration. The final outputs 
of this procedure are the obscurer parameters,
the obscured SED illuminating the WA,
and the rescaled WA ionization parameters.\\
%
%
%
%
%
%
At first, 
we dealt with the third
and longest \leg spectrum (Obs. CH3)
and we fitted it jointly
with the simultaneous
NuSTAR spectrum. We started with
a few iterations where
only the continuum 
was allowed to vary.
However, since
in this way the fit was left
with large residuals
(C/Expected C=846/307, Fig. \ref{chspettro.fig}, first panel)
we released
first the covering fractions
(Fig. \ref{chspettro.fig}, second panel)
and in turn the column densities
of both the obscurer phases
(Fig. \ref{chspettro.fig}, third panel).
Once we achieved the best fit,
we applied it
to the other two \leg spectra,
for comparison.
For obs. CH1,
the fit tends
to steepen the continuum
up to $\Gamma \sim 2.2$.
In contrast,
we could easily fit Obs. CH2 just
by renormalizing the model.
Therefore we decided 
that we can stack observations CH2 and CH3, 
and thus increase the signal-to-noise ratio 
of the spectrum.\\
We fitted this stacked
spectrum
(hereafter labeled
as Obs. CH2+3) 
together with NuSTAR
using the iterative procedure 
just described.
The final
best fit model
is shown
in Fig. \ref{chspettro.fig} 
(fourth panel) 
and the final obscured SED 
produced in the iterative
fitting is plotted in Fig.
\ref{sed.fig}, dashed line. 
Using the above best fit, 
we tested whether 
the difference in photon index 
for observation CH1 could be due 
to changing properties of the obscurer.
If we keep 
the continuum shape frozen, 
the best fit 
prefers zero covering fraction 
for the cold obscurer.
For
this dataset,
we favor
this solution
because a large
variation of the continuum slope
would be inconsistent with
what NuSTAR and INTEGRAL
have shown
for the whole campaign (see U15).\\
All the best fit parameters
for the \chandra observations
are shown in Table \ref{allpar.tab}.
Compared to the spectra of the core of
the campaign,
the \chandra spectra
require both a steeper continuum
($\Gamma \sim 1.8$) 
and a lower covering fraction
of the warm obscurer (\cvw $\sim 0.8$).
However, 
in principle,
it is possible that a change 
in the ionization state
of the obscurer mimics
a drop in the
covering fraction.
The data quality is not sufficient 
to fit the ionization parameter, 
therefore 
we refitted the spectrum
with the ionization parameter 
of the warm obscurer expected 
if it responds immediately to flux changes. 
We used the UVW2 flux to calculate 
the expected increase in ionization parameter. 
In the fit we used the 
covering fractions as given in K14.
With these constraints
the resulting fit is statistically
worse (C/Expected C=647/311)
and shows larger positive
residuals in the \leg
band. 
Leaving the ionization parameter 
of the warm obscurer 
free, a better fit is derived,
but the obtained ionization parameter
($\log \xi \sim 0.9$)
is unrealistically high 
($\sim$100 times than the
average value)
considering the increase 
by only a factor 2 
in the UVW2 flux
during the flare.
Therefore, we can exclude
that a change in the
ionization of the obscurer
is the dominant cause
of the observed flare.\\
During
the \xmm \th \th campaign,
the discrete features of the WA
are always blended
with the obscured continuum
and not detectable.
During the \chandra observation
the source was about twice
as bright, and some WA signatures 
are visible in the stacked CH2+3 spectrum
(Fig. \ref{lines.fig}).
These features are consistent
with the WA model computed 
via our iterative procedure.
The broad trough
at $\sim$ 16 \AA\  is a blend of
unresolved transition array (UTA) 
from several ionized iron species
Between
20 and 24 \AA\ 
some \ion{O}{iv}-\ion{O}{v} 
absorption lines
are present.
In table \ref{righe.tab},
we list which of 
the \ion{O}{iv} and \ion{O}{v}
lines predicted by our WA model
contribute to each
feature. 
The WA comprises 6 ionization
components that could be 
distinguished thanks
to the excellent data quality of the
2002 spectrum.
Here, the lower
statistics does not
allow to overcome the blending
among the components.
%
%
%
%


\begin{table}[t]
\caption{List of the \ion{O}{iv} and \ion{O}{v}
lines predicted by our WA model that contributes
to the features visible in Fig. \ref{lines.fig}.}
\centering
\begin{tabular}{lcccc}
\hline
Ion &
WA component &
Outflow velocity &
$\lambda_{\rm obs}$ \tablefootmark{a}&
$\tau$ \tablefootmark{b}\\
& & \kms & \AA & \\
\hline
\multirow{4}{*}{\ion{O}{v}} & C & 1148 & 20.23 & 27 \\
& B & 547 & 20.27 & 7\\
& D & 254 & 20.29 & 3\\
& E & 792 & 20.26 & 1\\
\hline
\multirow{2}{*}{\ion{O}{iv}} &  B & 547 & 21.04 & 3\\
&  C & 1148 & 21.00 & 2\\
\hline
\multirow{3}{*}{\ion{O}{v}}& C & 1148 & 22.67 & 143\\
& B & 547 & 22.71 & 34\\
& D & 254 & 22.73 & 16\\
\hline
\multirow{2}{*}{\ion{O}{iv}}& C & 1148 & 23.04 & 10\\
& B & 547 & 22.08 & 5\\
\hline
\end{tabular}
\tablefoot{
\tablefoottext{a}
{Predicted wavelength,
considering the cosmological redshift and the blueshift
due to the outflow.}
\tablefoottext{b}
{Line optical depth.}}
\label{righe.tab}
\end{table}



\begin{figure}[]
    \includegraphics[width=0.5\textwidth]{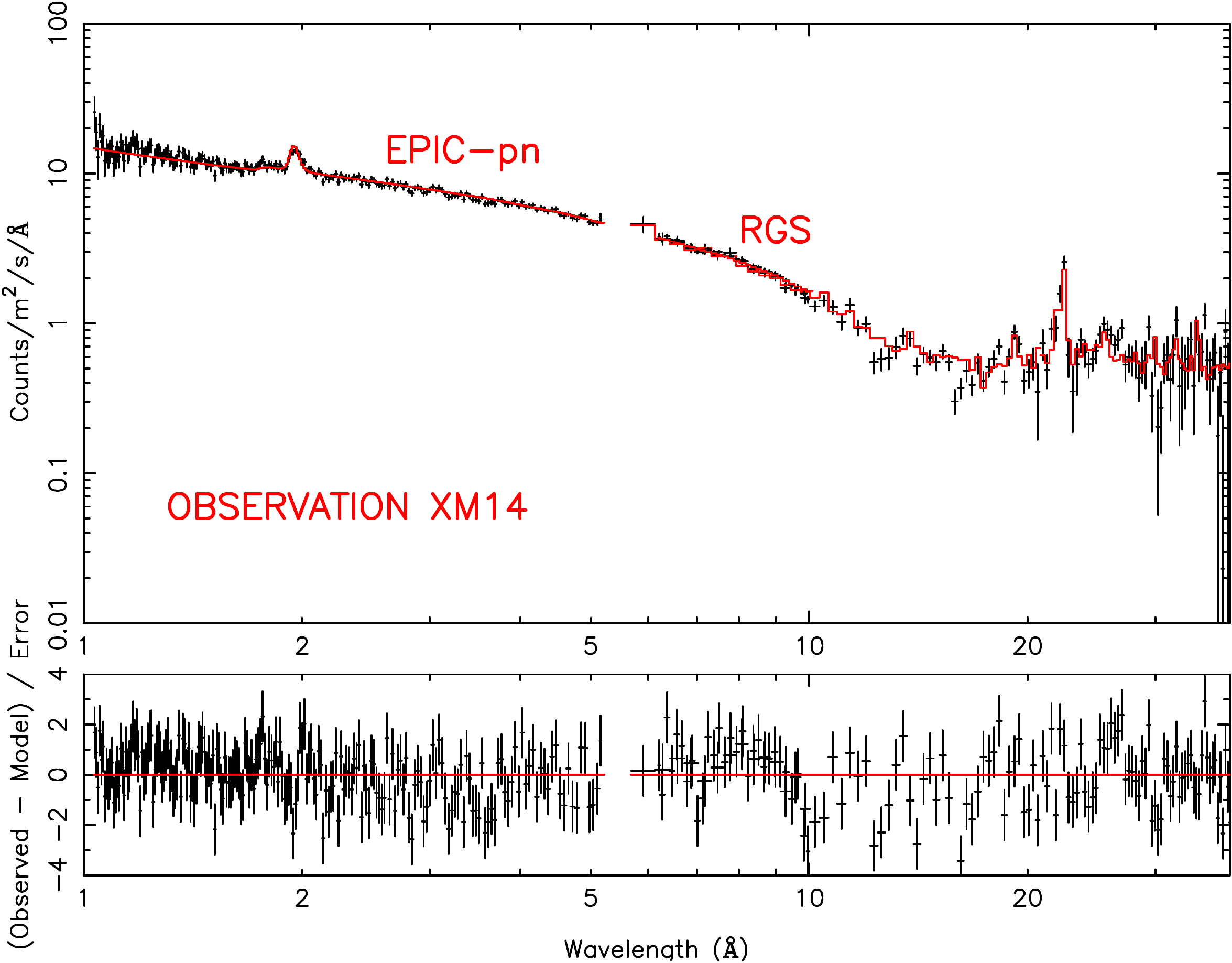}
  \caption{\xmm \th \th spectrum and model residuals for Obs. 14.
  The solid line represent our best fit model.
  The data have been rebinned for clarity.
  }
\label{xm14.fig}
\end{figure}


\subsection{The end of the campaign: 
the observations of December 2013 and February 2014}
\label{fit3}

In the last two
\xmm \th \th
observations of the campaign,
\sou was again at the same flux level 
of summer 2013.
Therefore, 
at first we attempted to fit them
using again the same WA
of the average spectrum.
The parameters of the
continuum and of the two
phases of the obscurer
were free.
This attempt 
resulted 
in a satisfactory fit for XM13,
while for Obs. XM14,
we obtained
a high column density
for the cold obscurer 
(\nhc$\sim 7 \times 10^{23}$ \colc) 
largely inconsistent 
with what was observed
throughout the campaign.
Thus,
we attempted to refit
this dataset
freezing the cold
column density
to the average value
measured 
in the previous observations.
This resulted
in negligible
covering fraction
for the cold obscurer
(\cvc$\la 10^{-2}$) and
in a flatter continuum
($\Gamma \sim 1.6$ instead of
$\sim$1.7).
This lower value of $\Gamma$
is also similar 
to the other values 
measured at this luminosity
(Fig. \ref{gacorr.fig}).
For these reasons,
we considered
this
solution more physically
plausible.
We note
that for this
dataset a solution
with a negligible covering fraction
of the cold obscurer and
a flatter continuum is found
also in U15 when including
in the fit INTEGRAL data at
higher energies. 
\\ 
%
%
%
Considering 
the large change in the obscurer, 
also for this observation 
we used our iterative method 
to determine the ionization of the WA 
(for details see Sect. \ref{chfits}).
%
%
%
%
%
We built
the SED for Obs. XM14 interpolating
the Comptonization model of Paper I
repeating all the steps
already described in Sect. \ref{chfits}. 
In this case, the slope of the continuum
at X-ray energies is provided
by the fit of the EPIC-pn data.
Starting from the solution
just described,
the fit converged
in a few iterations.
We also made
some iterations where
the ionization parameter
of the warm obscurer 
was left free. 
However,
since it was not possible
to constrain it,
we finally kept 
it frozen
to the value found in K14
(see Sect. \ref{templ}).
We adopt the final
result of 
the iterative routine
as our best-fit model.
The parameters
resulting from this exercise
are outlined in Table
\ref{allpar.tab}.\\
We show the final best-fit
model for Obs. XM14 in
Fig. \ref{xm14.fig}.
The final WA model
that we found is only
slightly different
from the average one.
The residuals
of the best-fit model
display some
possible features at $\sim$10.3 \AA\
and $\sim$12.3 \AA\, which
are close to the expected
wavelength of the 
\ion{Ne}{ix}-K edge
(10.54 \AA)
and \ion{Ne}{X}-\lia \th
(12.34 \AA) respectively.
We checked if adding
additional absorbing
column density in
\ion{Ne}{ix}-\ion{Ne}{x}
could improve the fit.
For this, we used the SLAB
model in SPEX, that allows
us to fit a single ionic 
column density regardless
of the ionization balance with the
other ions.
We found however that this
additional component 
is not required
by the fit ($\Delta C=-1$).


\section{Discussion}
\label{disc4}
\subsection{The short-term variability}
\label{dvar.sec}
\begin{figure*}[]
  \includegraphics[angle=180,width=1.0\textwidth]{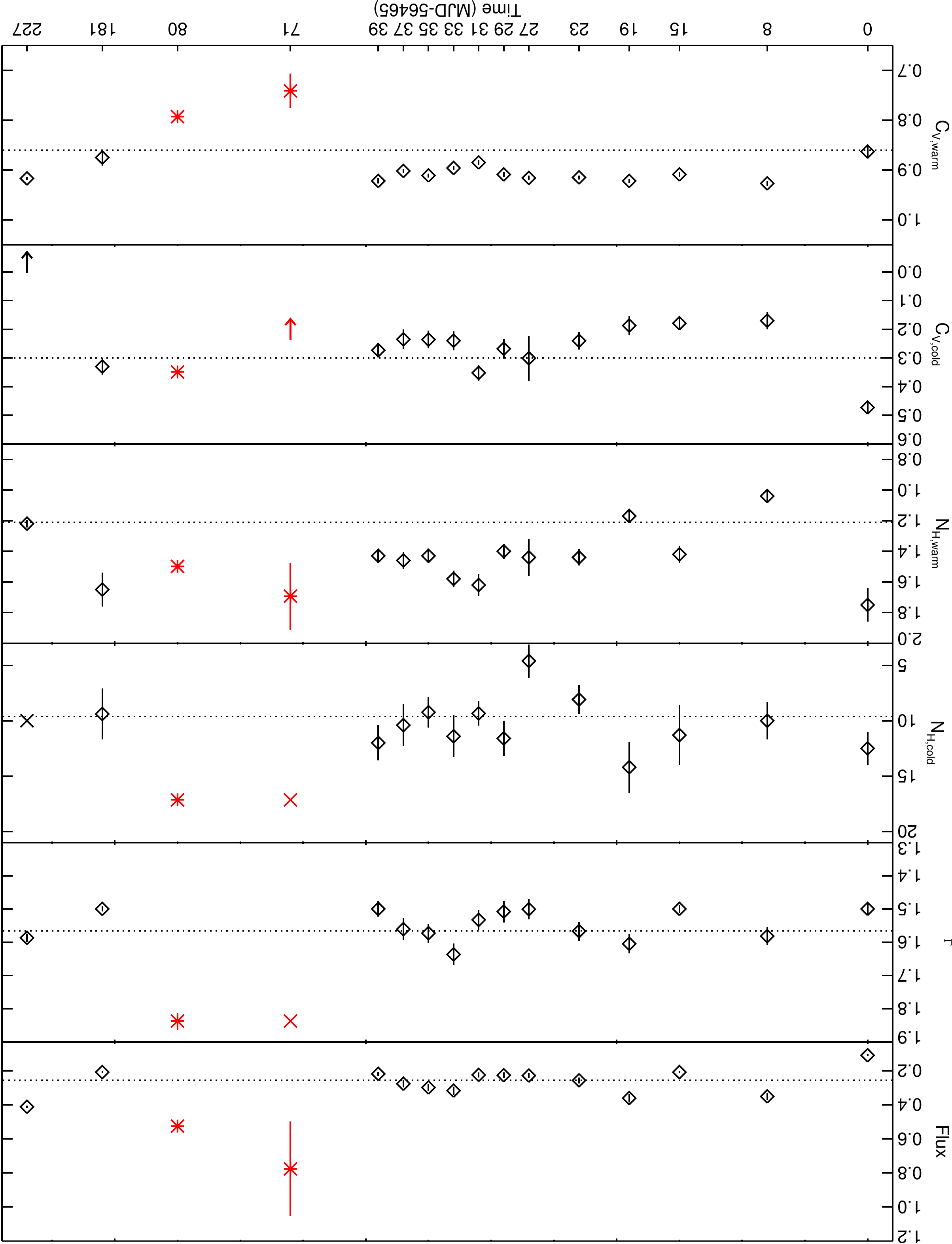}
  \caption{From the top to the bottom panel:
  the observed flux in the 0.3--2.0 keV band, 
  the photon index of the primary continuum,
  the column densities of the cold 
  and the warm obscurer, and the covering fractions of the
  cold and warm obscurer are shown as a function of time.
  The flux is plotted in units of $10^{-11}$ \ergsc. The column
  densities are plotted in units of $10^{22}$ \colc.
  The Modified Julian Day (MJD) correspondent to each observation 
  is labeled on the horizontal axes,
  which has been shrunk for display purpose. 
  Black diamonds and red asterisks
  identify parameters measured
  with \th \xmm \th and Chandra-LETGS, respectively. 
  Error bars, when larger than the size of the plotting symbol, are also shown.
  Upper limits are plotted as an arrow.
  Crosses represent values that were kept frozen in the fits.
  In each panel, the dotted horizontal line indicates the parameter value
  derived in K14 for the co-added XM1--12 spectrum.}
 \label{var.fig}
\end{figure*}

During our extensive monitoring campaign 
in 2013 and early 2014
NGC 5548 was always obscured.
In this analysis
we have applied
the model developed in K14 
for the average spectrum
of the core of the campaign
to all the individual observations,
with the aim of understanding 
how the source
varies.
When both the intrinsic continuum
and the obscurer are allowed to vary,
the model is able to 
explain the variability 
on the 
2 days--8 months
timescale sampled in the
monitoring campaign.
The obscuring
material that is causing the
soft X-ray flux depression of \sou
varies along our line of sight,
both in column density
and in covering fraction.
The scenario proposed by K14, 
that the source is obscured 
by a patchy wind, 
is consistent 
with our variability findings.
In this framework,
the variability of the obscuration 
may well be due to several reasons, 
e.g. motion across the line of sight 
and changing ionization 
with continuum variability.\\
The best fit parameters
for the continuum and
the two phases of the obscurer
are displayed 
in Fig.\ref{var.fig}
as a function of time.
During the core 
of the \th \xmm \th \th campaign
(Obs. XM1--XM12)
the source was steadily
obscured
and the variability in flux 
was relatively small
($\sim$ 27\%).
The only clear outlier 
with a flux significantly
different from the average
is Obs. XM1.
The values we found
for the warm obscurer
parameters
(Fig. \ref{var.fig}
4th and 6th panel)
deviate 
from those found
in K14 
the co-added XM1--XM12
spectrum. 
This is due to our 
different modeling of
the soft-excess, that
dominates the continuum
in the band where 
the absorption
from this component 
is more effective.
In our modeling
the Comptonized
soft-excess,
whose normalization
is set by the UV flux
measured by Swift for
each observation,
is always more luminous
($L^{\rm COMT}_{\rm 0.3-2.0 \, keV}
=[0.8-1.4] \times 10^{43}$ \ergs)
than the phenomenological
blackbody fitted in K14.
Throughout 
the core of the campaign,
the intrinsic continuum
is fairly constant in shape
(with a standard deviation
of the spectral index
$\sigma_{\rm \Gamma} \sim 3 \%$)
and slightly variable in normalization
($\sigma_{\rm UVW2} \sim 17\%$,
$\sigma_{\rm Norm} \sim 24\%$ 
for the soft-excess and the
power law component,
respectively).
For the obscurer, 
the cold component 
was the most variable
($\sigma_{\cvc} \sim 31 \%$
and $\sigma_{\nhc} \sim 23 \%$).
In contrast, 
for the warm component 
the covering fraction is stable 
($\sigma_{\cvw} \sim 2 \%$) 
and the column density 
shows rather small variability 
($\sigma_{\nhw} \sim 13\%$).
The large deviation
from the average of the
cold covering fraction 
suggests that
the obscurer inhomogeneity,
that is possibly 
dominated by the cold phase,
may have caused most of
the variability observed
during this phase 
of the campaign.\\
In this paper
we have also presented
the \leg datasets 
that were acquired 
in September 2013 when
\sou underwent a two-week
brightening. 
With respect
to the core of the campaign
changes in both 
the continuum
and the obscurer
are required to fit
these spectra.
At the time 
of the \chandra
observation, the UV flux
measured by Swift,
which in our interpretation
is a tracer 
for the soft X-ray excess,
was the highest of the whole
campaign. 
At the same time,
the continuum at hard X-ray 
energies
increased in flux
and became steeper.
For both observations 
the warm obscurer component 
has a significantly 
lower covering fraction.
In the first 
observation
also the covering fraction
of the cold component is lower.
As pointed out
in Sect. \ref{ch.sec},
a variation in the obscurer
ionization alone is 
insufficient to explain
to observed variation
in spectral shape.\\
The decrease
in covering fraction, 
in principle,
can be due
either to a local 
thin patch of the 
obscurer
passing in our line of
sight at the moment of the 
\chandra or 
to a geometrical 
change of the UV/soft 
X-ray continuum source
behind the obscurer.
The former possibility,
although it cannot be
excluded, would however
require that the intrinsic
continuum and the obscurer
change properties in
synchrony, which seems
\textit{ad hoc}.
In the Comptonization 
model of \citet{pet2013},  
the UV/soft X-ray spectrum 
is supposed to be produced 
via Comptonization
of the UV disk photons 
in a ``warm'' ($T \sim $ 1 keV) 
and moderately thick 
($\tau \sim 10-20$) 
corona. 
A ``hot'' corona, 
with higher temperature 
($\sim$ 100 keV) 
and smaller optical depth 
($\sim$ 1), 
will in turn Compton upscatter
these UV/soft X-ray photons  
to hard X-ray energies.
In this interpretation,  
an increase in physical size 
of the warm corona,
while naturally augmenting
both the UV and the soft X-ray flux,  
would also result in a drop 
of the observed 
obscurer covering fraction. 
Moreover, 
the increase 
in the UV/soft X-ray photon flux 
will more effectively cool the 
hot corona,  
hence producing a steeper 
hard X-ray spectrum 
which is in agreement with 
the observation.\\
%
%
%
%
%
Due to increased 
soft X-ray flux
in the
\chandra observations
some discrete 
WA features
(Fe UTA, \ion{O}{iv}--\ion{O}{v})
became evident in the spectrum.
These are 
the only detectable WA signatures
in any X-ray 
spectrum of our campaign.
These features
are best fitted by a
WA which has a 
significantly lower
degree of ionization
than what is observed 
in the unobscured 2002 spectrum.
Like K14, for all the WA components
we found best fit ionization parameters
which are 0.40 dex lower than 
the 2002 values ( $\log \xi_{A-E} ^{2002}$=
0.78, 1.51, 2.15, 2.36, 2.94, and 3.13). 
This means
that the ionizing luminosity received
by the WA decreased by a factor of 
$\sim$ 4. Thus, 
our analysis confirms the
K14 finding. The decrease
in the WA ionization,
that is seen also in the UV
\citep{ara2015}
is explained
when the newly discovered 
obscurer is located between
the nucleus and the WA.
In this geometry,
the obscurer shadows the
central source and
prevents most of the ionizing
flux from reaching 
the warm absorber.\\
%
%
%
%
%
%
The absorbers in \sou
changed again
in the last observation 
of the
campaign, namely Obs. XM14.
We found that 
in this dataset, 
the covering fraction
of the cold obscurer
became negligible.
At the same time,
the continuum above 2.0 keV
is similar to what is observed
throughout the \xmm \th \th 
campaign while the soft-excess
component is only slightly higher.
Thus,
the most likely cause of 
the spectral changes observed
in this dataset is again
the inhomogenity
of the obscuration.\\
Variability
in the continuum
and obscurer parameters
is also noticed in the
U15 analysis of the
six \xmm \th \th and
one \chandra observations
that were acquired
simultaneously
with a higher energy
observation.
However, the parameters
obtained fitting
the EPIC-pn
jointly with RGS
as done here
are not directly comparable
with those obtained
fitting the EPIC-pn
jointly with
NuSTAR and/or 
INTEGRAL, as done
in U15.
This is both
because of cross-calibration
issues between 
the instruments
and of differences
in the analysis.
In particular,
as noticed 
also in U15,
RGS and EPIC-pn 
are mismatched 
in flux in the 
overlapping band
as a function 
of energy
\citep[e.g.,][]{det2010}.
On the other hand, 
NuSTAR spectra
are systematically
steeper 
than EPIC-pn spectra 
($\Delta \Gamma \sim 0.1$
see Cappi et al., in prep).
In the present analysis
we consider also the ionized phase
of the obscurer (K14), 
while U15 use two purely neutral
components. This can
affect the broadband curvature
of the model. Moreover
U15 has an additional
degree of freedom in the
high-energy cutoff of the
continuum. 
For all these reasons,
the only meaningful comparison
is among the overall
trend of the parameters.
Even taking into account
the differences between 
our analysis and the one
presented in U15
the parameters
trends that we discuss
below still hold.\\ 
%
%


\begin{figure}[t]
 \includegraphics[angle=180, width=0.5\textwidth]{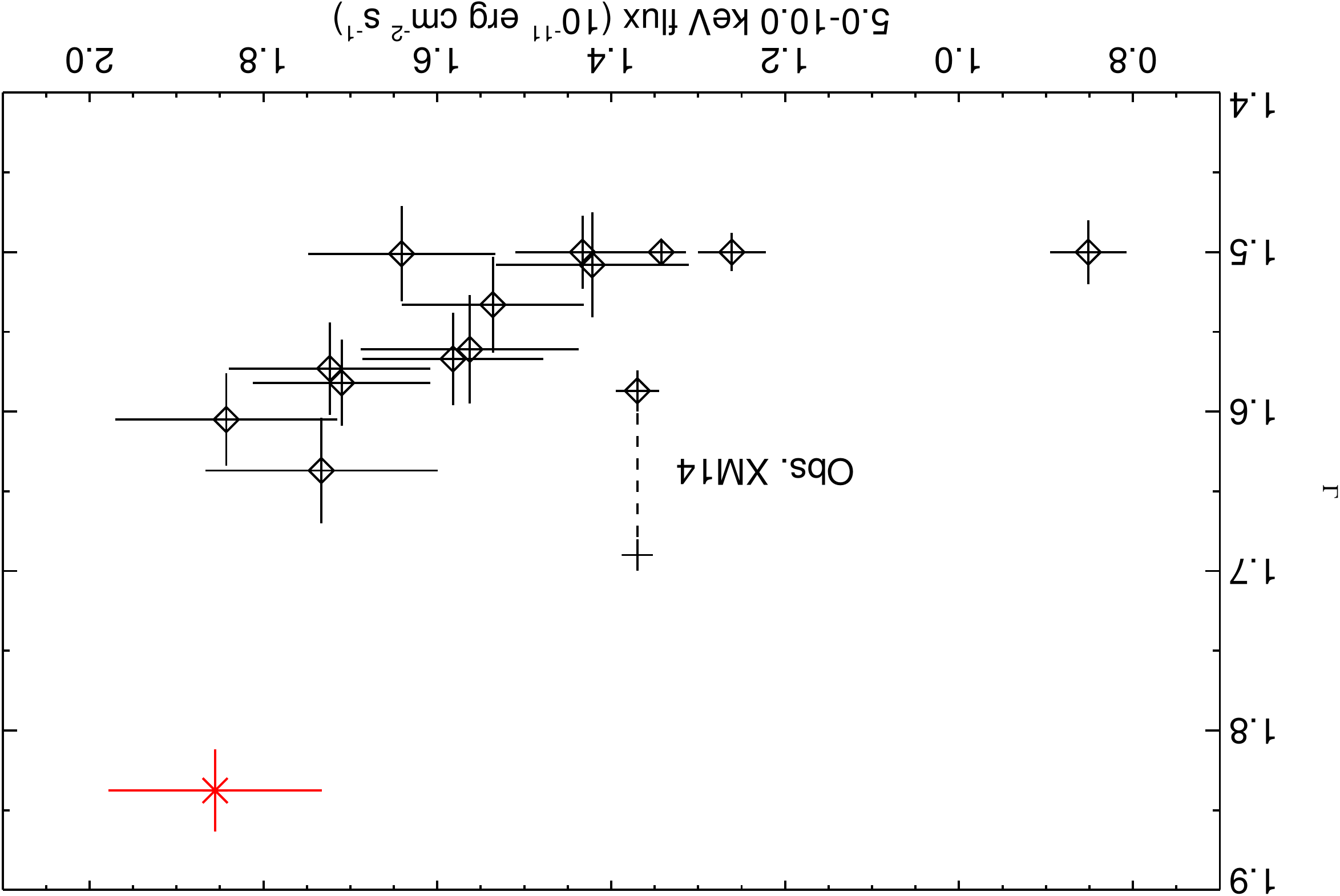}
  \caption{Fitted continuum slope as a function of the observed
   5.0--10.0 keV flux for all the \xmm \th \th and \chandra datasets, with errors.
   Black diamonds and red asterisk
   identify parameters measured
   with \th \xmm \th and Chandra-LETGS, respectively. 
   The black cross represents the higher value of $\Gamma$
   that, for Obs. XM14, would be required by a fit
   including the cold obscurer (see Sect. \ref{fit3}).}
\label{gacorr.fig}
\end{figure}

\subsection{What drives the variability?}
\label{corr.sec}
\begin{figure*}[t]
  \includegraphics[angle=180,width=1.0\textwidth]{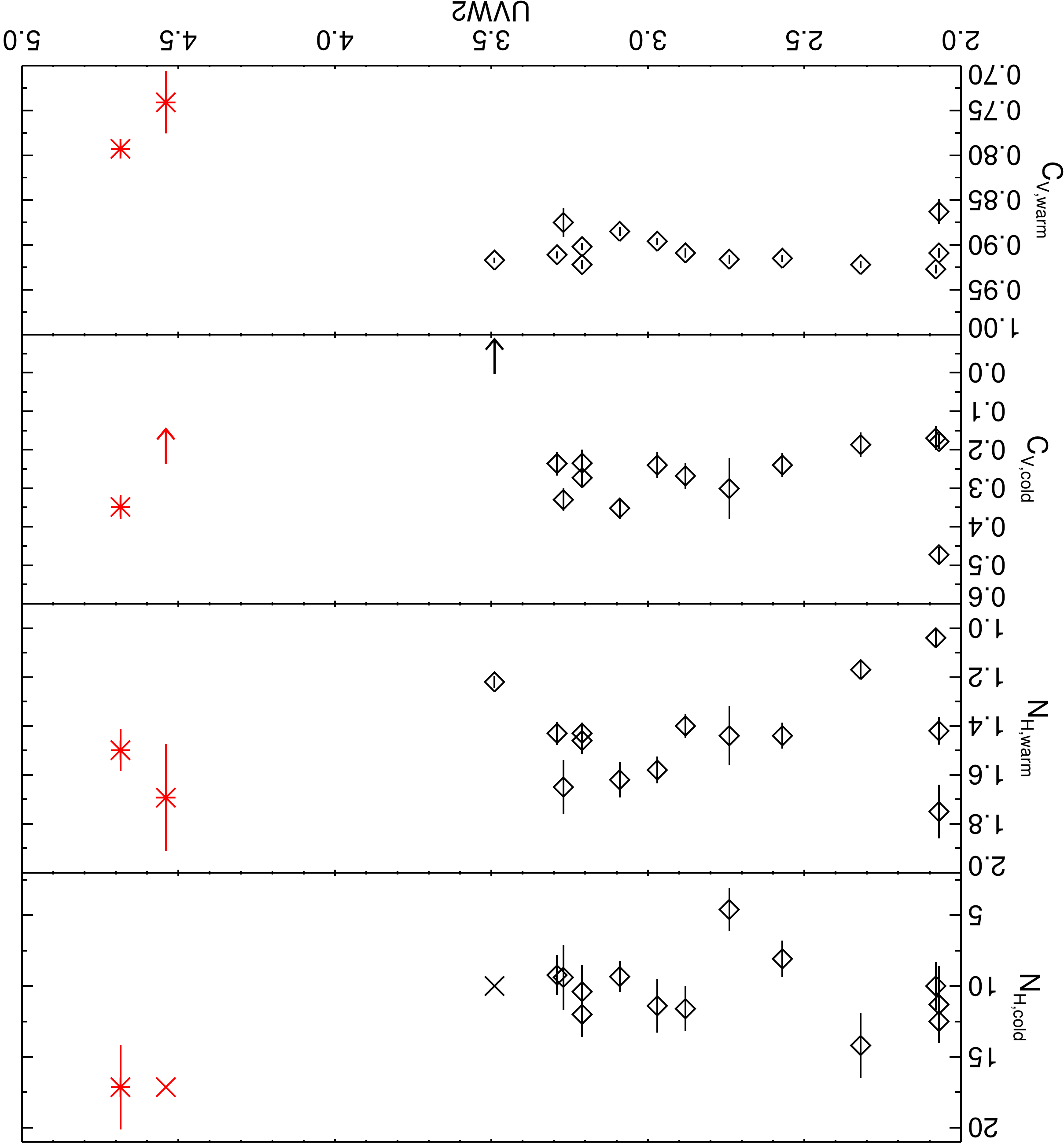}
  \caption{From the top to the bottom panel:  the column densities
  of the cold and the warm obscurer, the covering fraction of the cold
  and the warm obscurer are plotted as function of the flux
  in the UVW2 filter. Column densities are plotted in units of
  $10^{22}$ \colc. The UVW2 flux is plotted in units of
  $10^{-14}$ \ergsca.
   Black diamonds and red asterisks
  identify parameters measured
  with \th \xmm \th and Chandra-LETGS, respectively. 
  Error bars, when larger than the size of the plotting symbol, are also shown.
  Upper limits are plotted as an arrow.
  Crosses represent values that were kept frozen in the fits.}
\label{ocorr.fig}
\end{figure*}

To understand
if there are 
some systematic 
factors
driving the
short-term variability
of the source,
we looked
for correlations among
the best fit parameters
and the unobscured
flux measured 
for the 16 datasets
presented here.
We used the
hard X-ray flux in the
5.0--10.0 keV band
and the UVW2 flux
listed 
in Table \ref{obs4.tab}
as tracers of the
intrinsic continuum, 
as they are
almost unaffected 
by the obscuration.
Even considering these
unabsorbed bands,
the range of flux 
sampled
in the monitoring
campaign is narrow
(a factor of $\sim$2),
with the only outliers
at lower and higher
flux being Obs. XM1
and Obs CH2+3.
Therefore,
to evaluate
the reliability
of any correlations
we checked if 
it is still holding 
when removing these
two data points
from the computation
of the Pearson
correlation
coefficient.\\
In Fig. \ref{gacorr.fig}
we show that 
the best fit continuum slope
steepens as the hard
X-ray flux increases.
For the complete
sample the correlation
is extremely significant.
The Pearson correlation
coefficient is 
$R_{\rm all}=0.85$ 
implying a probability 
$p \sim 10^{-5}$
for the null hypothesis.
When excluding Obs. XM1
and Obs. CH2+3 from the
computation,
the degree of correlation
still remains significant
($R_{\rm XM2-XM14}$=0.68,
$p_{\rm XM2-XM14}$=1\%).
This trend
has been
already noticed in the past
for \sou in \citet{kaa2004}
and has been also reported
in other Seyfert galaxies
\citep[e.g., MCG 6-30-15][]{shi2002},
with different interpretations
\citep[see e.g.,][]{pon2006,gia2013}.
In the same Fig. \ref{gacorr.fig}
we show also that
for Obs. XM14,
the higher value of
$\Gamma$ that
would be required
by a fit including
a thick cold obscurer
(that we rejected,
see. Sect. \ref{fit3})
is inconsistent with
the correlation
valid for all 
the other datasets.\\
In Fig. \ref{ocorr.fig}
we plot the parameters
of both the obscurer
phases as function
of the UVW2 flux. 
The only parameter
showing a possible trend
with the intrinsic 
continuum is the
warm covering factor.
Namely, the drop 
observed during the 
\chandra observations
may be the tail
of a mild decreasing
trend visible also for
the \xmm \th \th data
points (Fig. \ref{ocorr.fig},
bottom panel). 
Formally, when the two \chandra data points
are included in the computation
a significant correlation
($R_{\rm all}=-0.73$,
$p_{\rm all}$=1\%) is present.
When considering only the
\xmm \th \th sample,
the trend is only qualitative
($R_{\rm XM1-XM14}=-0.03$
$p_{\rm XM1-XM14}$=25\%).
In Sect. \ref{dvar.sec}
we have suggested that the
drop in covering fraction 
observed during the September 2013
flare is due to an increase
in the size of the 
soft X-ray/UV source.
A clear correlation between
the warm covering fraction and intrinsic
continuum, supported
by more numerous data points
at different flux values,
would favor the hypothesis
that this is a systematic
effect producing
at least part of the
observed covering fraction
variability.
This
trend is not apparent
for the cold covering fraction.
This could be due to
a higher degree of inhomogenity
in the cold phase that would
also explain its larger
variability in covering fraction
(e.g., it went 
from 0.47 in
Obs XM1 to 0
in Obs. XM14).\\
In conclusion,
a combination
of changes
in the continuum
and in the obscurer
physical parameters
is required
to explain the short-term
spectral variability
of NGC 5548 during
our 2013-2014 campaign.
The lack of correlation 
between the intrinsic continuum
(as traced by the UVW2 flux)
and obscurer parameters 
indicate that the obscurer
must physically 
change properties 
independent on the 
source flux level.
The case of the September 2013 spectrum
suggests that the soft X-ray 
emitting region change geometry 
as the flux increases.
This could be a systematic effect
contributing to the overall covering
fraction variability.


%
\section{Summary and conclusions}
\label{conc4}

During the multiwavelength monitoring
campaign that we performed in 2013-2014
for the Seyfert 1 galaxy NGC 5548,
the source
had a soft X-ray flux well below
the long-term average,
except for a two-week long flare
in September 2013.
In K14,
we have ascribed this condition
to the onset of a persistent,
weakly-ionized but high-velocity 
wind that blocks $\sim$90\% 
of the soft X-ray flux
and lowers the ionizing luminosity
received by the WA.
Thus,
in this condition,
the normal WA
that was previously
observed in this source
is still present, but with
a lower ionization. 
In this paper
we fitted all 
the high-resolution
\xmm \th \th and \chandra
datasets that were taken 
during the campaign with a model
that consistently accounts
for a variable continuum,
the newly discovered
obscurer and the new
ionization conditions
of the historical WA.
We found that:
\begin{enumerate}
 \item On the timescales sampled in the monitoring campaign
       (2 days--8 months) both the intrinsic continuum
       and the obscurer are variable. The obscuring material
       varies both in column density and in covering fraction. 
       This rapid variability is consistent with the
       picture of a patchy wind proposed by K14.
       
\item  The \chandra spectra that were taken
       just after the peak of the flare in September 2013
       are explained by both an increase and a steepening 
       of the intrinsic continuum and a drop 
       in the obscurer covering fraction.
       The latter is likely to be due to a geometrical change
       of the soft X-ray continuum source behind the obscurer.
       
\item  The \chandra spectra of September 2013
       show absorption from Fe-UTA, \ion{O}{iv} and \ion{O}{v},
       consistent with belonging to the
       lower-ionized counterpart of the historical
       \sou warm absorber. These 
       are the only individual WA features
       in any X-ray spectrum of the campaign.
 
\item  A positive 
       correlation between the X-ray continuum
       slope and the observed 5.0--10.0 keV 
       flux  holds for both 
       the \xmm \th and \chandra
       datasets. 

\item The addition of the two \chandra points
      produce a formal anticorrelation
      between the warm obscurer covering fraction
      and the intrinsic continuum luminosity,
      as traced by the observed UVW2 flux.


\end{enumerate}

%

\begin{acknowledgements}


This work is based on observations obtained with \xmm \th \th
an ESA science mission with instruments and contributions directly
funded by ESA Member States and the USA (NASA). 
This research has made
use of data obtained with the NuSTAR mission, a project led by the California
Institute of Technology (Caltech), managed by the Jet Propulsion Laboratory
(JPL) and funded by NASA. 
%
%
This work made use of data supplied by the UK Swift Science
Data Centre at the University of Leicester. 
We thank the Chandra team for allocating
the LETGS triggered observations.
We thank the International Space Science
Institute (ISSI) in Bern for their support and hospitality. 
SRON is supported
financially by NWO, the Netherlands Organization for Scientific Research. 
%
%
M.M.acknowledges support from NWO and the UK STFC. 
This work was supported
by NASA through grants for HST program number 13184 from the Space Telescope
Science Institute, which is operated by the Association of Universities
for Research in Astronomy, Incorporated, under NASA contract NAS5-26555.
M.C. acknowledges financial support from contracts ASI/INAF n.I/037/12/0 and
PRIN INAF 2011 and 2012. 
P.O.P and F.U. acknowledge funding support 
from the CNES and the french-italian International Project 
of Scientific Collaboration: PICS-INAF project n°181542
%
%
%
S.B. and A.D.R. acknowledge INAF/PICS 
financial support and financial support from the
Italian Space Agency under grant ASI/INAF I/037/12.
A.D.R acknowledge financial support from contract PRIN INAF 2011.
%
%
G.P. acknowledges support via an EU Marie Curie
Intra-European fellowship under contract no. FP-PEOPLE-2012-IEF-331095
and Bundesministerium für Wirtschaft und Technologie/Deutsches Zentrum für
Luft- und Raumfahrt (BMWI/DLR, FKZ 50 OR 1408). 
FU acknowledges support from 
Universit\'e Franco-Italienne (Vinci PhD
fellowship).
M.W. acknowledges
the support of a PhD studentship awarded by the UK STFC. 
%

\end{acknowledgements}




\end{document}